\documentclass[nolineno]{jfm}

\usepackage{graphicx}
\usepackage{newtxtext}
\usepackage{newtxmath}
\usepackage{natbib}
\usepackage{hyperref}
\hypersetup{
    colorlinks = true,
    urlcolor   = blue,
    citecolor  = black,
}

\newcommand{\RomanNumeralCaps}[1]
\linenumbers

\renewcommand{\vec}[1]{\boldsymbol{#1}}

\newcommand{\bmth}[1]{\mbox{\boldmath $#1$}}
\newcommand{\grad}{\bmth{\nabla}}
\newcommand{\Fr}{\textrm{Fr}}
\newcommand{\St}{\textrm{St}}

\title{Bulldozing an immersed granular material in a confined channel}

\author{Liam C.~Morrow\aff{1,2}   \corresp{\email{liam.morrow@eng.ox.ac.uk}},
  Oliver W.~Paulin\aff{1,3},
  Matthew G.~Hennessy\aff{4},
  Duncan R.~Hewitt\aff{5},
  Miles L.~Morgan \aff{6},
  Bjørnar Sandnes \aff{6},
  \and Christopher W.~MacMinn~\aff{1}\corresp{\email{christopher.macminn@eng.ox.ac.uk}}
 }

\affiliation{\aff{1} Department of Engineering Science, University of Oxford, Oxford OX1 3PJ, United Kingdom
\aff{2} Research School of Earth Sciences, The Australian National University, Acton ACT 0200, Australia
\aff{3} Max Planck Institute for Dynamics and Self-Organization, Göttingen 37077, Germany
\aff{4} School of Engineering Mathematics and Technology, University of Bristol, Bristol, BS8 1TW, United Kingdom
\aff{5} Department of Applied Mathematics and Theoretical Physics, University of Cambridge, Cambridge, CB3 OWA, United Kingdom
\aff{6} Department of Chemical Engineering, Swansea University, Swansea, SA1 8EN, United Kingdom}

\begin{document}
\maketitle

\begin{abstract}
The motion of an immersed granular material in a channel is characterised by complex interactions among the grains, between the grains and the permeating liquid, and between the grains and the channel walls. Here, we develop a reduced-order continuum model for the bulldozing of an immersed, sedimented granular material by a piston in a channel. In our continuum approach, the granular layer and the overlying fluid layer evolve as a system of coupled thin films. We model the granular phase as a dense, porous, visco-plastic material that experiences Coulomb-like friction with the walls. Conservation of mass and momentum under a thin-film approximation lead to an elliptic equation for the velocity of the grains that is coupled with an evolution equation for the height of the granular layer. We solve our model numerically for a variety of different scenarios to explore the interactions between wall friction, internal viscous-like stresses, and fluid flow above and through the granular layer. We complement our numerical results with a series of experiments that provide insight into the validity and limitations of the model.
\end{abstract}

\section{Introduction} \label{sec:Introduction}

Depending on the confining stress, granular materials can respond mechanically at the continuum scale as a solid, liquid, or gas \citep{Forterre2008}. Granular flows are relevant to a wide variety of natural processes, including avalanches \citep{Doppler2007,Pailha2009}, terrestrial and submarine landslides \citep{Hampton1996,Legros2002}, blood flows through small vessels \citep{Freund2011}, and debris flows \citep{Iverson1997}, as well as to a variety of industrial applications. The continuum rheology of granular materials has been studied extensively, primarily focusing on internal friction and thus on relative motion and sliding \textit{within} the granular material. For example, many studies consider scenarios in which a granular material is confined and sheared between a fixed plate or cylinder and a moving one \citep{Aussillous2013, Houssais2016, Maurin2016, Ouriemi2009}, whereas classical bedload transport involves the shearing of an unconfined granular material by an overlying fluid flow \citep{Frey2011}. Avalanche-type flows have also attracted substantial attention, in which an unconfined granular material flows downhill under gravity \citep{Bouchut2016, Cassar2005, Doppler2007, Forterre2008, Pailha2009}. Finally, \citet{Sauret2014} studied the unconfined pile-up of a granular material during bulldozing by a vertical plate on a rotating table, focusing on the initial growth of the granular pile and then on its long-time shape.

Additional challenges arise in confined granular-flow problems where the grains are able to slide along the confining walls. The flow of sand through an hourglass is a relatable example; similar interactions occur within subsurface faults and fractures \citep{Dorostkar2017} and in silos \citep{Janssen1895, Landry2003}. This class of problems has received less attention. Some authors have considered the impact of the confining vertical side walls on avalanche-type flows in channels \citep{Jop2005, Taberlet2008, Artoni2015}. Simulations and experiments show that these walls can have a significant impact on the flow, potentially changing the convexity of the flow profile if the channel is sufficiently narrow. \citet{Marks2015} used the discrete-element method (DEM) to simulate the piston-driven bulldozing of dry grains between horizontal top and bottom plates in two spatial dimensions. They found that, once the grains have piled up to bridge the gap between the two plates, the bulldozing force grows exponentially with the downstream length of the bridged region due to sliding friction between the grains and the plates, which is consistent with the classical Janssen effect \citep{Janssen1895}. \citet{Morgan2020} performed experiments to study the motion of granular material along a confined, quasi-2D channel where the motion of the grains is driven by a net flow of liquid through the channel. They observed that the velocity profile of the grains was non-monotonic in the vertical direction.

Additional inspiration for the present study comes from a series of experiments involving the bulldozing of grains by capillary forces. This problem has been considered in both one- and two-dimensional geometries. In the one-dimensional case, air was injected into a capillary tube containing a wet granular suspension \citep{Dumazer2016,Dumazer2020}. Because the air is non-wetting to the grains, the liquid--air meniscus acts to bulldoze the grains like a piston. When the grains pile-up to bridge the gap, the granular plug becomes jammed and the gas phase eventually percolates through the static plug and emerges on the other side, resulting in a repeating pattern of granular plugs alternating with gas regions. In the two-dimensional case of a Hele-Shaw cell, the system has an additional degree of freedom \citep{Eriksen2018,Sandnes2007,Sandnes2011,Zhang2023}. The gas always pushes the granular material in the direction of least resistance, resulting in the growth of a fingering pattern with a morphology that is distinct from the traditional viscous fingers that would form in the absence of the grains \citep{Paterson1981, Saffman1958}. It has been shown that ad hoc, rule-based simulations can reproduce these patterns \citep{Eriksen2015,Eriksen2015b,Knudsen2008}, but a true quasi-1D or quasi-2D continuum model derived from the fundamental mechanics of the problem is still lacking.

\begin{figure}
	\centering
	\includegraphics[width=0.6\linewidth]{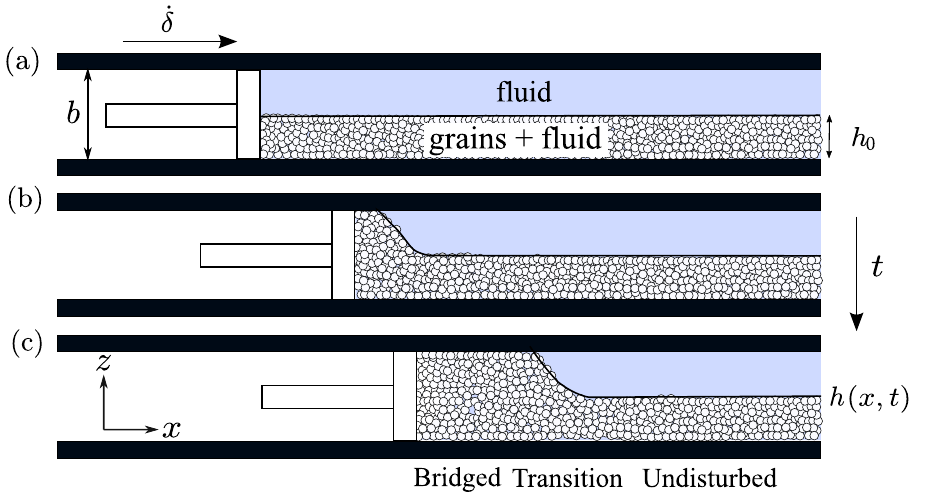}
	\caption{A schematic of granular bulldozing in a confined, horizontal channel. (a)~The system consists of two parallel plates separated by a gap $b$. The gap is initially filled with a mixture of fluid and grains, with the grains forming a uniform, fluid-saturated layer of height $h_0$ on the bottom plate. (b)~A piston moves from left to right with constant velocity $\dot{\delta}$, bulldozing both grains and fluid. The grains pile up near the piston and eventually bridge the gap between the plates. (c)~As bulldozing continues, the domain comprises three regions: a bridged region where $h = b$, a wedge-shaped transition region where $h_0 < h < b$, and an undisturbed region where $h = h_0$ and the solid is stationary. \label{fig:schematic} }
\end{figure}

Here, we develop a continuum model to describe the bulldozing of an immersed granular material in a confined, horizontal geometry. We then use this model to explore the interactions between fluid flow, granular motion and pile-up, and sliding friction. To do so, we focus on the scenario illustrated in Figure~\ref{fig:schematic}, where a thin gap between two parallel plates is filled with solid grains and a fluid. We take the grains to be denser than the fluid, such that they sediment in a uniform layer on the bottom plate. A piston then bulldozes both grains and fluid from left to right at a constant velocity. This problem is a generalisation of the work by \citet{Marks2015}, who considered the limiting case where the fluid phase is ignored. We formulate a two-phase, two-layer model comprising a single Newtonian fluid in the top layer and an immersed granular material (grains and fluid) in the bottom layer, such that depth-averaging leads to a system of coupled thin-film flows. For the granular phase, we adopt the established $\mu(I)$-type rheology \citep{Jop2005,Pouliquen2006,Cassar2005}, where the granular continuum behaves like a yield-stress fluid with a pressure-dependent yield stress and a rate-dependent friction coefficient $\mu$. We consider both the inertial $\mu(I)$ rheology \citep{Midi2004,DaCruz2005,Jop2005} and the viscous $\mu(I_v)$ rheology \citep{Cassar2005,Boyer2011}, which are distinguished by whether the viscosity of the interstitial fluid is expected to play an unimportant role or a dominant role in the mechanics, respectively. Further, we impose simple Coulomb-like friction at the top and bottom walls, such that the grains will only slide along the walls if the local shear stress is sufficiently large relative to the local normal stress \citep{Pennestri2016}.

The remainder of this work is organised as follows. In \S\ref{sec:ModelDevelopment}, we derive our model based on the approach described above. As previously mentioned, the interactions between the grains and the walls are governed by Coulomb friction. This model gives an expression for the shear stress at the walls only if the magnitude of the shear stress is sufficiently large, which is not known \textit{a priori}. To address this difficulty, we introduce a novel regularisation that gives an expression for the shear stress over the entire computational domain. In \S\ref{sec:DryRegion} and \S\ref{sec:ViscousModel}, we present a numerical study of our model. In \S\ref{sec:DryRegion}, we consider the inertial (`dry') $\mu(I)$ rheology where the viscosity of the fluid phase is negligible. We investigate the granular velocity profile in the region where the grains have bridged the gap, as well as in the wedge-shaped transition region (see fig.~\ref{fig:schematic}b,c). In \S\ref{sec:ViscousModel} and \S\ref{sec:Experiments}, we consider the viscous (`wet') $\mu(I_v)$ rheology and focus on the influence of the fluid on the grains, comparing results from our model with laboratory experiments. Finally, in \S\ref{sec:Discussion}, we summarise our findings, comment on the validity of our modelling assumptions, and suggest avenues for future work.

\section{Model development} \label{sec:ModelDevelopment}

We consider the piston-driven bulldozing of a dense, fluid-saturated granular material in a confined horizontal channel. The channel consists of a gap of height $b$ between two parallel, horizontal plates. The gap is filled with an incompressible Newtonian fluid and a granular material with mean grain diameter $d$. The solid grains are negatively buoyant and form a layer of height $h(x, t)$ on the bottom plate. The granular layer has a uniform initial height $h(x,0)=h_0$ [figure~\ref{fig:schematic}a]. A rigid piston located at $x = \delta(t)$ travels from left to right (\textit{i.e.}, in the positive $x$-direction) at a constant speed $\dot{\delta}$, bulldozing both grains and fluid [figure~\ref{fig:schematic}(b) and (c)]. Note that we ignore motion in the $y$-direction (\textit{i.e.}, into the page).

We denote the local volume fractions of the fluid and solid phases as $\phi_f$ and $\phi_s$, respectively, such that $\phi_f + \phi_s = 1$. For simplicity, we take $\phi_s$ to be uniform and equal to a constant $\phi_s^*$ within the granular layer and zero outside the granular layer. Thus, we neglect elastic compression of the grains and shear-induced compaction and dilation of the granular layer in order to focus on the physical interactions of internal friction, pile-up, wall friction, and fluid flow. The solid-fraction distribution is then given by
\begin{align}
	\phi_s(x, z, t) = 
	\begin{cases} 
		\phi_s^*, & \textrm{for } 0 \le z \le h \\
		0,            & \textrm{for } h < z \le b
	\end{cases}.
\end{align}
We denote the fluid velocity by $\vec{v}_f = u_f(x, z, t) \vec{e}_x + w_f(x, z, t) \vec{e}_z$ and the solid velocity by $\vec{v}_s = u_s(x, z, t) \vec{e}_x + w_s(x, z, t) \vec{e}_z$, with $\vec{e}_x$ and $\vec{e}_z$ the unit vectors in the $x$- and $z$-directions, respectively. The volume-averaged velocity is then $\vec{v} = u(x, z, t) \vec{e}_x + w(x, z, t) \vec{e}_z = \phi_f \vec{v}_f + \phi_s \vec{v}_s$.  Conservation of mass for the solid phase can then be written in vertically integrated form as
\begin{align}
	\frac{\p h}{\p t} + \frac{\p}{\p x} (\bar{u}_{s} h) = 0, \label{eq:hEvolution}
\end{align}
where 
\begin{align}
	\bar{u}_{s} = \frac{1}{h} \int_{0}^{h} u_{s} \, \textrm{d} z
\end{align}
is the average horizontal velocity of the solid. Similarly, conservation of total volume requires that the average horizontal velocity on any vertical slice must be equal to the piston speed,
\begin{align}
	 \frac{1}{b} \int_{0}^{b} u \, \textrm{d} z = \dot{\delta}. \label{eq:Incompressible2}
\end{align}

Before discussing the model in detail, we briefly consider the anticipated structure of the flow, as sketched in figure~\ref{fig:schematic}. We expect the flow to be predominantly horizontal and quasi-static, provided, as discussed in \S\ref{sec:Nondimensional} below, that the macroscopic Froude number and fluid Reynolds number are sufficiently small that friction and viscosity dominate the momentum balance. As noted above, the granular layer has a frictional yield strength. Thus, sufficiently far ahead of the piston, the grains must be undisturbed at their original height $h_0$. Nonetheless, the motion of the piston dictates that there must be a net rightward flow of material at all values of $x$ (eq.~\ref{eq:Incompressible2}); in the far field, we therefore expect a net fluid flow both through and above the grains. Near the piston, we expect grains to pile up and `bridge' the gap (figure~\ref{fig:schematic}b), leading to a translating `transition' region that links the bridged near-piston region with the undisturbed far-field region (figure~\ref{fig:schematic}c). Once formed, the bridged region must grow steadily in length as grains pile up ahead of the piston. The transition region can then either grow continually in length (if its leading edge travels faster than its trailing edge, both of which must travel faster than the piston) or evolve toward a steady shape that translates like a travelling wave (if its leading and trailing edges travel at the same speed, again faster than the piston). In the latter scenario, conservation of solid volume dictates that the transition region must travel at a speed $b \dot\delta/(b-h_0) > \dot\delta$. We observe both of these scenarios in the results below. An alternative scenario, also observed in some cases (\S\ref{sec:ViscousModel}), is that the grains fail to displace all of the viscous fluid from the overlying fluid layer, and thus never fully bridge the gap. In this scenario, there is no bridged region and the transition region must grow continually.

\subsection{Mechanics}\label{ss:mechanics}

We focus on the limit in which the channel is sufficiently thin, or the piston sufficiently slow, that the inertia of the fluid--solid mixture is unimportant in the macroscopic momentum balance. In this limit, which we quantify more formally in \S\ref{sec:Nondimensional}, conservation of linear momentum can be written
\begin{align}
	\nabla \cdot \vec{\sigma} = \rho g \vec{e}_z,  \label{eq:ConservationMomentum}
\end{align}
where $\vec{\sigma}$ is the total stress in the mixture and $\rho = \phi_f \rho_f + \rho_s \phi_s$ is the local volume-averaged density of the mixture, and where we have adopted the convention that tensile stresses are positive. Going forward, we assume that the ratio of typical vertical to horizontal length scales is sufficiently small that we can adopt a lubrication-type approximation, in which $w \ll u$ and $\p_x \ll \p_z$. We discuss the validity of these assumptions in \S\ref{sec:Experiments} and \S\ref{sec:Discussion}.

\subsubsection{Upper layer (fluid)} \label{sec:UpperLayer} 

The upper layer, $h<z<b$, comprises a single incompressible Newtonian fluid of viscosity $\eta_f$, for which the lubrication approximation suggests that the fluid pressure $p_f$ is hydrostatic, $p_f(x,z,t) = P_f(x,t) + \rho_f g (b - z)$, such that
\begin{align}
	\frac{\p \sigma_{xz}}{\p z} = \frac{\p P_f}{\p x} \qquad \textrm{with} \qquad \sigma_{xz} = \eta_f \frac{\p u_f}{\p z},
\end{align}
where $P_f$ is the unknown fluid pressure at $z = b$. We impose a no-slip boundary condition on the flow at the top wall, $u_f = 0$ on $z = b$, as is standard. The appropriate boundary condition at the interface $z = h$ is less obvious, but, for simplicity, we follow previous models of bedload transport and impose no tangential slip between the overlying fluid and the granular solid by requiring that $u_f = u_s$ at $z=h^+$ \citep[\textit{e.g.},][]{Aussillous2013}. Note that this choice does not constrain the fluid velocity on the granular side of the interface, where in general we expect a nonzero Darcy flow (\textit{i.e.}, fluid flow relative to the solid grains), and neither does it prevent fluid flow across the interface. Note also that, in reality, some degree of tangential slip is expected at the boundary between an external fluid flow and a porous medium \citep[\textit{e.g.},][]{LeBarsWorster2006}. Many associated slip boundary conditions have been proposed, most famously the so-called Beavers--Joseph condition \citep{Beavers1967, LeBarsWorster2006}. We revisit this point briefly below, at the end of \S\ref{sec:LowerLayer}.

Imposing these two no-slip conditions, we arrive at an expression for the velocity of the fluid in the upper layer:
\begin{align}
	u_f(x, z, t) = \frac{(b - z)(h - z)}{2 \eta_f} \frac{\p P_f}{\p x} + \left(\frac{b - z}{b - h}\right)u_s(z = h).\label{eq:fluidVelocity}
\end{align}
The shear stress exerted on the granular layer by this flow is then 
\begin{align}
	\sigma_{xz}(z = h) = \eta_f \left[ \frac{h - b}{2 \eta_f} \frac{\p P_f}{\p x} - \frac{u_{s}(z = h)}{b - h} \right]. \label{eq:FluidShear}
\end{align}
Equation \eqref{eq:fluidVelocity} suggests that the flow in the upper layer is a Poiseuille--Couette flow, combining a parabolic profile from the background pressure gradient with a linear shear profile from the motion of the granular layer relative to the top wall.

\subsubsection{Lower layer (fluid--solid mixture)} \label{sec:LowerLayer}

For the lower layer, we adopt Terzaghi's principle \citep{Terzaghi1936} by decomposing the total stress into the fluid pressure and the effective solid stress $\vec{\sigma}'$,
\begin{align}
	\vec{\sigma} = \vec{\sigma}' - p_f \vec{I}.  \label{eq:Terzaghi}
\end{align}
Following \citet{Jop2006}, we then decompose $\vec{\sigma}'$ into its isotropic and deviatoric components,
\begin{align}
	\vec{\sigma}' = -p_s \vec{I} + \vec{\tau},  \label{eq:LinearStress}
\end{align}
where $p_s = -(1/3) \textrm{tr}(\vec{\sigma}')$ and $\textrm{tr}(\vec{\tau}) = 0$. Further, we impose that the fluid flows through the porous, granular material according to Darcy's law,
\begin{align}
	(1 - \phi_s^*) (\vec{v}_f - \vec{v}_s) = -\frac{k}{\eta_f} \left( \grad p_f - \rho_f g \vec{e}_z \right), \label{eq:Darcy}
\end{align}
where $k$ is the permeability of the granular skeleton, which we expect to scale with the square of the grain size: $k\sim{}d^2$. 

Combining equations~\eqref{eq:ConservationMomentum}, \eqref{eq:Terzaghi}, and \eqref{eq:LinearStress} under the lubrication approximation, we arrive at
\begin{align}
	\frac{\p}{\p z} (p_s + p_f) = -\rho g \qquad \textrm{and} \qquad \frac{\p \sigma_{xz}'}{\p z} = \frac{\p}{\p x} (p_s + p_f), \label{eq:abc123}
\end{align}
along with $\sigma_{xx}' = \sigma_{zz}' = -p_s$. The vertical component of Equation~\eqref{eq:Darcy} suggests that the fluid pressure is hydrostatic; further requiring that it be continuous across $z=h$ leads to $p_f = P_f + \rho_f g (b - z)$.  It then follows that $p_s = P_s(x, t) + \phi_s^* \Delta \rho g (h - z)$, where $\Delta \rho = \rho_s - \rho_f$ and $P_s$ is the as-yet-unspecified solid pressure at the interface $z = h$. As such,
\begin{align}
	\sigma_{xz}' = \sigma_{xz} &= \sigma_{xz}(z = h) - \left( h-z \right)  \left( \frac{\p P}{\p x} + \phi_s^* \Delta \rho  g \frac{\p h}{\p x}  \right), \label{eq:sigmaXZ}
\end{align}
where $P = P_f + P_s$. Further, the horizontal component of equation~\eqref{eq:Darcy} becomes
\begin{align}
	(1 - \phi_s^*) (u_f - u_s) = - \frac{k}{\eta_f} \frac{\p P_f}{\p x}.
    \label{eq:Darcy2}
\end{align}
Note that continuity of normal stress requires that $P_s$ must vanish when the solid does not bridge the gap (\textit{i.e.}, when $h < b$), such that $P_s \ne 0$ only in the bridged region (\textit{i.e.}, where $h \equiv b$). Correspondingly, conservation of mass requires that the fluid and solid in the bridged region move at the same speed, and thus $\partial_x P_f = 0$ in the bridged region. 

Imposing no tangential slip between the granular solid and the fluid in the upper layer, as we did above, implies a jump in tangential fluid velocity between the free flow in the upper layer and the Darcy flow in the lower layer (\textit{i.e.}, across the interface). In reality, we anticipate a narrow transition region at the interface (on the scale of one grain/pore, or $\sim{}d\sim \sqrt{k}$) over which the fluid velocity smoothly adjusts \citep[see, \textit{e.g.},][]{Beavers1967, LeBarsWorster2006}. We expect the difference between applying a no-slip condition and a more elaborate slip condition to scale with $\sqrt{k}$, which is typically extremely small. As noted above, this boundary condition does not constrain flow \textit{across} the interface; rather, the normal component of the fluid velocity is dictated by conservation of mass, as is standard in lubrication problems.

Note that the incomplete model above already provides some insight into the flow in the far field, where the grains are undisturbed at uniform depth $h_0$ and the fluid flows both above them and through them. For stationary grains ($u_s=0$), the flow in the upper layer is a simple Poiseuille flow (eq.~\ref{eq:fluidVelocity}) and the uniform Darcy flow through the lower layer follows from eq.~(\ref{eq:Darcy2}). The total flow rate is $\dot{\delta}b$, as determined by the piston, with a fraction $(b-h_0)^3/[(b-h_0)^3+12 h_0 k]$ flowing in the upper layer and a fraction $12 h_0 k/[(b-h_0)^3+12 h_0 k]$ flowing through the grains in the lower layer.

\subsubsection{Granular friction and wall friction} \label{sec:GranularFriction}

As noted above, we adopt a $\mu(I)$-type rheology for the internal mechanics of the granular material \citep{Jop2005,Pouliquen2006,Cassar2005}. In these model constructions, the internal mechanical interactions between grains are taken to obey a Coulomb-like friction model,
\begin{alignat}{2}
	|\sigma_{xz}'| &= \mu p_s \quad &\textrm{if } \left| \sigma_{xz}' \right|  \ge \mu_s p_s, \label{eq:CoulombMuS} \\
	\frac{\p u_s}{\p z} &= 0 &\textrm{otherwise},  \label{eq:CoulombMuS2}
\end{alignat}
where $\mu$ is a dimensionless, rate-dependent coefficient of kinetic friction, as discussed further in \S\ref{sec:Rheology}, and $\mu_s$ is a dimensionless constant coefficient of static friction. If the magnitude of the shear stress is sufficiently small, equation \eqref{eq:CoulombMuS2} indicates that grains will resist internal rearrangement and the solid phase will act locally as a rigid body. If the shear stress is sufficiently large, the solid phase will yield such that $|\p_z u_s| > 0$ and the solid velocity will satisfy equation~\eqref{eq:CoulombMuS}, with the sign of $\sigma_{xz}'$ given by the sign of the shear rate $\p_z u_s$. As a result, the vertical profile $u_s(z)$ may feature multiple coupled yielded and un-yielded regions separated by free boundaries. We consider the various possible flow scenarios in \S\ref{sss:velocity_profile} below. We specify the granular rheology in more detail in \S\ref{sec:Rheology}.

At the bottom surface of the granular layer, we assume that the mechanical interactions between the grains and the wall obey a Coulomb-like friction model:
\begin{align}
	\sigma_{xz}'(z = 0) \le \mu_w p_s(z = 0), \label{eq:ShearFrictionBot}
\end{align}
such that
\begin{align}
	u_s(z = 0) = 0  \quad &\textrm{if } \quad\sigma_{xz}'(z = 0) < \mu_w p_s(z = 0), \label{eq:CoulombMuW2} \\
	u_s(z = 0) > 0  \quad &\textrm{if } \quad\sigma_{xz}'(z = 0) = \mu_w p_s(z = 0), \label{eq:CoulombMuW1}
\end{align}
where $\mu_w$ is the dimensionless constant coefficient of friction between the grains and the wall. Note that we assume that $u_{s} \ge 0$, such that $\sigma_{xz}'(z=0)\ge0$ and, thus, that friction resists relative motion. To generate a single, continuous problem in $x$, we unify these conditions in a regularised form in \S\ref{sec:Nondimensional}.

The appropriate boundary condition at the top surface of the granular layer depends on whether or not the grains are in contact with the top wall. If the grains have bridged the gap (\textit{i.e.}, if $h \equiv b$), then we similarly apply Coulomb friction at the top wall: 
 \begin{align}
	\sigma_{xz}'(z=b) \ge -\mu_w p_s(z=b) \quad\mathrm{if}\quad h\equiv{}b, \label{eq:ShearFrictionTop}
\end{align}
such that
\begin{align}
	u_s(z=h) = 0  \quad &\textrm{if } \quad\sigma_{xz}'(z=b) > -\mu_w p_s(z=b) \quad\mathrm{and}\quad h\equiv{}b, \label{eq:CoulombMuW2_top} \\
	u_s(z=h) > 0  \quad &\textrm{if } \quad\sigma_{xz}'(z=b) = -\mu_w p_s(z=b) \quad\mathrm{and}\quad h\equiv{}b, \label{eq:CoulombMuW1_top}
\end{align}
where the differences in sign relative to equations~\eqref{eq:ShearFrictionBot}--\eqref{eq:CoulombMuW1} reflect the different direction of the normal vector, again ensuring that friction resists relative motion for $u_s\geq0$. If the grains have not bridged the gap between the plates (\textit{i.e.}, if $h<b$), then we instead impose that the shear stress is continuous with the upper fluid layer via equation~\eqref{eq:FluidShear}. To generate a single, continuous problem in $x$, we also unify these conditions in a regularised form in \S\ref{sec:Nondimensional}.

\subsubsection{Velocity profile}\label{sss:velocity_profile}

\begin{figure}
	\centering
	\includegraphics[width=1\linewidth]{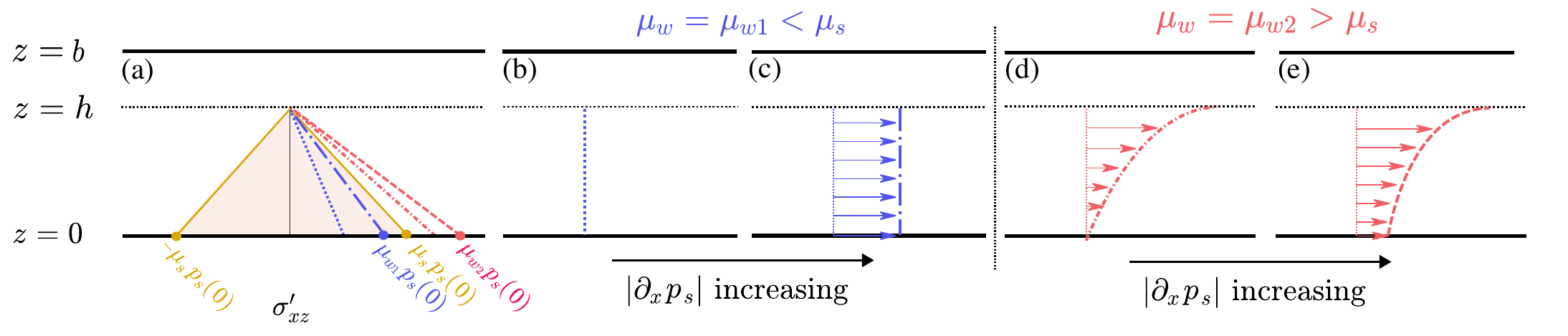}
	\caption{Examples of the different possible stress and velocity configurations in the unbridged state ($h < b$ and $P_s\equiv{}0$), neglecting the fluid phase ($\p_x{p_f}=0$ and $\sigma'_{xz}(z=h) = 0$). (a) Possible shear-stress profiles $\sigma_{xz}'(z)$ (red and blue lines), which must be linear from eq.~(\ref{eq:sigmaXZ}), together with the effective yield stress of the granular layer $\pm \mu_s p_s$, where $p_s = \phi_s^* g \Delta \rho (h - z)$ (solid yellow); the shear stress lies below the yield stress in the shaded region. Different colours correspond to whether the wall friction is less than the internal static friction, $\mu_w = \mu_{w1} < \mu_s$ (blue), or greater, $\mu_w = \mu_{w2} > \mu_s$ (red), with corresponding velocity profiles shown in (b,c) and (d,e), respectively. In each case, there are two possibilities: the shear stress on the bottom wall does not reach the slip stress, $|\sigma_{xz}'(z=0)|<\mu_w p_s(0)$, and the grains do not slip there (b,d), or it does and they do (c,e). Where the shear stress is below the internal yield stress, $|\sigma_{xz}'|<\mu_sp_s$, the granular layer behaves as a rigid plug (b,c).}
	\label{fig:sketchshear}
\end{figure}

\begin{figure}
	\centering
	\includegraphics[width=\linewidth]{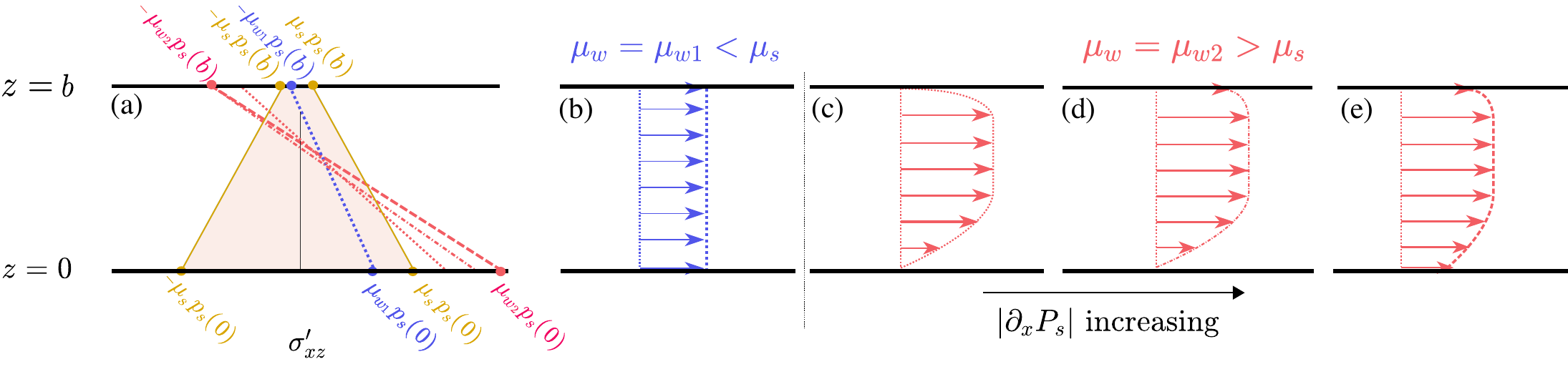}
	\caption{Examples of the different possible stress and velocity configurations in the bridged state ($h \equiv b$, $P_s>0$, and $\sigma'_{xz}(z=h) < 0$), neglecting the fluid phase ($\p_x{p_f}=0$). (a)~Possible shear-stress profiles $\sigma_{xz}'(z)$ (red and blue lines), together with the effective yield stress of the granular layer $\pm \mu_s p_s$, where $p_s = P_s + \phi_s^* g \Delta \rho (b-z)$ (solid yellow); the shear stress lies below the yield stress in the shaded region. If $\mu_w = \mu_{w1} < \mu_s$ (blue), the granular layer does not yield anywhere and must move as a rigid plug (b). If $\mu_w = \mu_{w1} > \mu_s$ (red), different flow profiles are possible depending on whether the shear stress attains the slip stress on either the upper or lower wall, as sketched in (c,d,e), each corresponding to the equivalent line style in (a).}
	\label{fig:sketchshearbridge}
\end{figure}

Before giving specific details of the granular rheology, we first present in figures~\ref{fig:sketchshear} and \ref{fig:sketchshearbridge} a selection of example velocity profiles that can occur in the transition zone ($h<b$) and in the bridged region ($h\equiv{}b)$, respectively. In both cases, the qualitative nature of the profile depends centrally on whether the coefficient of wall friction, $\mu_w$, is larger or smaller than the static coefficient of internal friction, $\mu_s$. 

We first consider the unbridged portions of the granular layer (\textit{i.e.}, the transition region and the far field) in figure~\ref{fig:sketchshear}, for which $h < b$ and $P_s \equiv 0$. For simplicity, we neglect the role of the fluid by taking $\sigma_{xz}'(z = h) = 0$ and $\p_xp_f=0$. Recalling that the shear stress at any horizontal location increases linearly with depth (eq.~\ref{eq:sigmaXZ}), we find that four types of qualitative profile arise. When $\mu_w < \mu_s$ (blue curves), the shear stress can never grow large enough for the granular layer to yield internally: $|\sigma_{xz}'| < \mu_s p_s$ and $|\p_z u_s| \equiv 0$ for all $z$. As a result, the grains do not deform and either remain stationary    (figure~\ref{fig:sketchshear}b) or move as a rigid body (figure~\ref{fig:sketchshear}c), depending on whether $\sigma_{xz}'(z = 0)$ is less than, or equal to, $\mu_w p_s(z=0)$, respectively.  By contrast, when $\mu_w > \mu_s$ (red curves), the granular layer yields over its entire thickness, again with either no slip (figure~\ref{fig:sketchshear}d) or slip (figure~\ref{fig:sketchshear}e) on the base, depending on the size of the basal stress. Note that analogous profiles have been observed in the case of viscoplastic fluids with wall slip \citep{Lawal1993}.

We next consider the bridged portions of the granular layer, in figure~\ref{fig:sketchshearbridge}, for which $h\equiv0$, $P_s>0$, and $\sigma_{xz}'(z=h= b)$ is dictated by friction with the top wall. Note that we again neglect the role of the fluid. As in the unbridged portions (figure~\ref{fig:sketchshear}), there are different cases depending on whether $\mu_w < \mu_s$ or $\mu_w > \mu_s$. In the former case (blue, figure~\ref{fig:sketchshearbridge}b), the grains remain everywhere unyielded and slide over both boundaries as a rigid plug (because of the moving piston, the bridged region cannot remain stationary). If $\mu_w > \mu_s$ (red, figure~\ref{fig:sketchshearbridge}c,d,e), there are more possibilities. The grains may be able to slip, or not, on both the upper and lower walls depending on the local shear and normal stresses, and in all cases there may be some unyielded interior region where the shear stress falls below the effective yield stress (figure~\ref{fig:sketchshearbridge}a). Across the entire parameter space, we expect the solid velocity to be smaller near the bottom wall than near the top wall because $p_s$ increases from top to bottom due to gravity.

To accommodate the scenarios shown in figures~\ref{fig:sketchshear} and \ref{fig:sketchshearbridge}, we seek in general a piecewise solid velocity of the form
\begin{align}
	u_s=\begin{cases}
		u_{s1}(x,z,t), & \text{if}\quad 0\le z \le h_{y1},\\
		u_{s2}(x,t), & \text{if}\quad h_{y1} < z \le h_{y2}\\
		u_{s3}(x,z,t), & \text{if}\quad h_{y2} < z \le h,\\
	\end{cases}, \label{eq:PiecewiseVelocity}
\end{align}
such that the granular layer is unyielded ($\p_z u_s = 0$) in an interior region $h_{y1} < z < h_{y2}$ and yielded ($|\p_z u_{s}| > 0$) outside of that region, with continuity of velocity requiring that $u_{s1} = u_{s2}$ at $z = h_{y1}$ and $u_{s2} = u_{s3}$ at $z = h_{y2}$. The locations $h_{y1}$ and $h_{y2}$ are determined by solving $\sigma_{xz}' = \pm \mu_s p_s$ to give
\begin{align}
	h_{y1} = h - \frac{\sigma_{xz}'(z = h) - \mu_s P_s}{\p_x P + \phi_s^* \Delta \rho g (\p_x h + \mu_s) } \quad \textrm{and} \quad
	h_{y2} = h - \frac{\sigma_{xz}'(z = h) + \mu_s P_s}{\p_x P + \phi_s^* \Delta \rho g (\p_x h - \mu_s) }.
\end{align}
In the yielded regions ($0 < z < h_{y1}$ and $h_{y2} < z < h $), we combine equations~\eqref{eq:sigmaXZ} and \eqref{eq:CoulombMuS} with the appropriate granular rheology from \S\ref{sec:Rheology} to obtain ordinary differential equations (ODEs) for $u_{s1}$ and $u_{s3}$.

\subsection{Granular Rheology} \label{sec:Rheology}

In the region where the granular layer has yielded, $|\sigma_{xz}'| > \mu_s p_s$, we adopt a $\mu(I)$-type rheology to describe its evolution~\citep{Jop2005,Pouliquen2006,Cassar2005}. These rheologies relate the effective viscosity of the solid phase, $\mu$, to either the inertial number, $I$, or the viscous number, $I_v$, both of which are defined below. The specific choice depends on the magnitude of the Stokes number, $\St \sim I^2/I_v 
\sim \rho_s d^2 |\p_z u_{s}|/ \eta_f$, which compares inertial forces to viscous forces on the scale of the grains. Viscous drag from the fluid is expected to dominate inertia at the grain scale for $\St\ll{}1$, whereas grain-scale inertia is expected to be dominant for $\St\gg{}1$. As such, different rheologies are appropriate in these two regimes. Some effort has been made to combine the two regimes into a single rheology \citep{Boyer2011,Baumgarten2019,Rettinger2022,Tapia2022}, but we treat them separately for simplicity.

In \S\ref{sec:DryRegion}, we consider the dry case where $\St\gg{}1$, in which limit the fluid phase is negligible. In this case, we follow \citet{Iordanoff2004} and \citet{DaCruz2005} by taking 
\begin{align}
	\mu(I) = \mu_s + \left(\frac{\mu_2 - \mu_s}{I + I_0}\right) I,  \label{eq:muI}
\end{align}
where $I_0$ and $\mu_2$ are constants. The inertial number $I$ is 
\begin{align}
	I = \sqrt{\frac{d^2 \rho_s}{p_s}} \left| \frac{\p u_s}{\p z} \right|, \label{eq:I}
\end{align}
where, under the lubrication approximation, $\left|\p u_s/\p z\right|$ is the leading-order contribution to the magnitude of the rate-of-strain tensor. Equation~\eqref{eq:muI} is an empirical fit to experiments and numerical simulations, and is defined such that $\mu\to\mu_s$ for $I\ll I_0$ and $\mu\to\mu_2$ for $I\gg{}I_0$.

In \S\ref{sec:ViscousModel}, we consider the wet case where $\St\ll{}1$, in which limit fluid viscosity dominates granular inertia. In this case, we follow \citet{Boyer2011} by taking
\begin{align}
	\mu(I_v) = \mu_s + \left(\frac{\mu_2 - \mu_s}{I_v + I_0}
    \right)I_v + I_v + \frac{5}{2}\phi_s^*\sqrt{I_v},  \label{eq:FullMuI}
\end{align}
where $I_0$ and $\mu_2$ are again constants, the appropriate values of which are not necessarily the same as in equation~\eqref{eq:muI}. The viscous number $I_v$ is
\begin{align}
	I_v = \frac{\eta_f}{p_s} \left|  \frac{\p u_s}{\p z} \right| , \label{eq:Iv}
\end{align}
having again taken $|\p_z u_s|$ to be the leading-order contribution to the magnitude of the rate-of-strain tensor. The last two terms on the right-hand side of equation~\eqref{eq:FullMuI} mean that $\mu(I_v)$ will increase without bound as $I_v$ increases.

By substituting either equations~\eqref{eq:muI} and \eqref{eq:I} or equations~\eqref{eq:FullMuI} and \eqref{eq:Iv} into the frictional-rheology model in equation~\eqref{eq:CoulombMuS}, we arrive at a non-linear ODE for $u_s$ with boundary conditions described in more detail below.

Note that both the $\mu(I)$ and the $\mu(I_\nu)$ models presented here have been shown to be mathematically ill-posed in their general, time-dependent, tensorial forms \citep{Barker_wellposed1,Barker_wellposed_susp}; specifically, they can be unstable to arbitrarily small-wavelength perturbations for some ranges of $I$ and $I_\nu$. This problem can be resolved by carefully accounting for compressibility in each case \citep{barker_wellposed_comp,Barker_wellposed_susp}. In the present lubrication framework, the momentum balance (eq.~\ref{eq:abc123}) is quasi-steady, which avoids any direct issues of ill-posedness. In the future, it could be instructive to consider a compressible version of the present model that remains formally well-posed even away from the lubrication limit.

\subsection{Non-dimensionalisation and regularisation} \label{sec:Nondimensional}

To non-dimensionalise our model, we use $b$, $b/\dot{\delta}$, $\dot{\delta}$, and $\phi_s^* \Delta \rho g b$ as characteristic spatial, temporal, velocity, and pressure scales. The key non-dimensional parameters are then
\begin{align}
	\theta = \frac{\eta_f \dot{\delta}}{b^2 \phi_s^* \Delta \rho g}, \quad \quad K = \frac{k}{b^2}, \quad \quad \textrm{and} \quad \quad\Fr = \frac{\dot{\delta} d}{b \sqrt{b g}},
    \label{eq:defs1}
\end{align}
as well as $\mu_w$, $\mu_s$, and the other constants in the $\mu(I)$ and $\mu(I_v)$ constitutive laws. The parameter $K$ is the non-dimensional permeability. Noting that $\eta_f \dot{\delta} / b$ is the characteristic shear stress from the upper fluid layer, the parameter $\theta$ can be viewed as a modified Shields parameter that measures the strength of fluid shear relative to the weight of the granular bed. The modified Froude number $\Fr$ measures the strength of inertial forces relative to gravitational forces, scaled by $d/b$. The ratio $\Fr^2/\theta \equiv \St$ recovers the Stokes number, which, as noted above, compares the relative importance of inertial and viscous forces on the scale of the grains. In the `wet' granular limit ($\St \ll 1$), we find that $\Fr$ drops out of the problem; in the `dry' granular limit ($\St \gg 1$), $\Fr$ becomes the main control parameter. 

In \S\ref{ss:mechanics} above, we neglected the inertia of the fluid--solid mixture in the macroscopic momentum balance (eq.~\ref{eq:ConservationMomentum}). The present scalings suggest that the relative importance of macroscopic inertia is governed by the square of the macroscopic Froude number, $\mathcal{F}^2 \equiv \dot\delta^2/(bg) = (\Fr \,b/d)^2 = \St \, \theta (b/d)^2$. Since the grains are much smaller than the gap, $\mathcal{F}$ is always larger than $\Fr$, meaning that neglecting macroscopic inertia is reasonable only when $\Fr^2 \ll 1$. Nonetheless, grain-scale inertia will control the granular rheology when $\St = \Fr^2/\theta \gg 1$ (\textit{i.e.}, in the `dry' limit). Together, these criteria suggest that macroscopic inertia can be neglected, even when the local dynamics are controlled by grain-scale inertia, when $\theta \ll \Fr^2 \ll 1$. In the `wet' limit (\textit{i.e.}, $\St\ll{}1$), $\Fr$ drops out and can be freely chosen to ensure that inertial effects are negligible. Note that we have also neglected the inertia of the overlying fluid layer, which is reasonable when the fluid Reynolds number $\mathrm{Re}_f = \rho_f \dot\delta b/\eta_f$ is sufficiently small.

From this point on, we retain our original notation but all quantities are dimensionless unless explicitly noted otherwise. The evolution equation for the height of the granular layer, equation~\eqref{eq:hEvolution}, becomes
\begin{align}
	\frac{\p h}{\p t} + \frac{\p}{\p x} (\bar{u}_s h) = 0\quad \textrm{with} \quad \bar{u}_s h = \int_0^{h_{y1}} u_{s1} \, dz + 
    (h_{y2} - h_{y1}) u_{s2} +
    \int_{h_{y2}}^h u_{s3} \, dz
   \label{eq:EvolutionH}
\end{align}
{and
\begin{align}
	h_{y1} = h - \frac{\sigma_{xz}'(z = h) - \mu_s P_s}{\p_x (P + h) + \mu_s } \qquad \textrm{and} \qquad	h_{y2} = h - \frac{\sigma_{xz}'(z = h) + \mu_s P_s}{\p_x (P + h) - \mu_s }, \label{eq:hy}
\end{align}
recalling that $P = P_f + P_s$, where $P_f$ and $P_s$ are the dimensionless fluid pressure at $z=1$ and the dimensionless solid pressure at $z=h$, respectively. To solve for the velocity profile $u_s(z)$, we equate equation~\eqref{eq:sigmaXZ} for the shear stress with equation~\eqref{eq:CoulombMuS} for the rheology to obtain
\begin{alignat}{2}
	\sigma_{xz}'(z=h) - (h - z) \left( \frac{\p P}{\p x} + \frac{\p h}{\p x} \right) &= \mu p_s, \qquad &0 \le z < h_{y1}, \label{eq:usNondim1} \\
	\frac{\p u_s}{\p z} &= 0, \qquad &h_{y1} \le z < h_{y2},  \label{eq:usNondim2} \\
	\sigma_{xz}'(z=h) - (h - z) \left( \frac{\p P}{\p x} + \frac{\p h}{\p x} \right) &= -\mu p_s \qquad &h_{y2} \le z < h,  \label{eq:usNondim3}
\end{alignat}
where $p_s = P_s + (h-z)$ is the dimensionless solid pressure and noting that the velocity profile $u_s(z)$ must be continuous in $z$. The shear stress at the top of the granular layer, $\sigma_{xz}'(z=h)$, is given below. As discussed in \S\ref{sec:Rheology}, we use a frictional rheology model with $\mu = \mu(I)$ for $\St \gg 1$ (\ref{eq:muI}) and $\mu = \mu(I_v)$ for $\St \ll 1$ (\ref{eq:FullMuI}), where the inertial and viscous numbers are
\begin{align}
	I = \frac{\Fr}{\sqrt{p_s}} \left|\frac{\p u_s}{\p z} \right| \quad \quad \textrm{and} \quad \quad I_v = \frac{\theta}{p_s} \left|\frac{\p u_{s}}{\p z}\right|, \label{eq:InertialNumber}
\end{align}
respectively, and the associated expressions for $\mu$ are given in equations~\eqref{eq:muI} and \eqref{eq:FullMuI}.

The shear stress $\sigma_{xz}^\prime(z=0)$ is determined by wall friction via equations~\eqref{eq:CoulombMuW2} and \eqref{eq:CoulombMuW1}. Similarly to internal granular friction, a challenge with wall friction is that the horizontal profile $u_s(x,z=0)$ may feature multiple coupled sliding and non-sliding regions. To avoid the need to track these regions explicitly, we unify equations~\eqref{eq:CoulombMuW2} and \eqref{eq:CoulombMuW1} in a dimensionless, regularised form as
\begin{align}
	\sigma_{xz}'(z=0) = \mu_w(P_s+h)\tanh \left[ \frac{u_s(z=0)}{\varepsilon} \right], \label{eq:RegularisedShearz=0}
\end{align}
where $\varepsilon \ll 1$ is a dimensionless regularisation parameter. Equation~\eqref{eq:RegularisedShearz=0} provides a single continuous expression for $\sigma_{xz}'(z = 0)$ over the entire $x$-domain, ensuring that $|\sigma_{xz}'(z = 0)| < \mu_w(P_s+h)$ when $u_s(z = 0) \lesssim{} \varepsilon \ll 1$ while converging rapidly to equation~\eqref{eq:CoulombMuW1} for larger values of $u_s(z=0)$.} This type of regularisation is common in the numerical simulation of viscoplastic fluids, because it avoids the need to introduce internal free boundaries between yielded and un-yielded regions~\cite[\textit{e.g.},][]{viscop}. Adopting it here for wall friction allows us to solve a single, continuous problem in $x$, which is convenient for the quasi-1D flow studied here and even more-so for complex, quasi-2D frictional-flow patterns~\citep[\textit{i.e.},][]{Sandnes2011}. Some care should be taken with regularisations of this form to avoid the potential misidentification of yielded regions~\citep[\textit{e.g.},][]{putz}; we illustrate the convergence of our solutions with decreasing $\varepsilon$ in Appendix~\ref{sec:regularisation}.

The shear stress $\sigma_{xz}^\prime(z=h)$ is determined locally by wall friction if the grains are in contact with the top wall (\textit{i.e.}, eqs.~\ref{eq:CoulombMuW2_top} and \ref{eq:CoulombMuW2_top} for $h\equiv{}1$), or by fluid shear otherwise (\textit{i.e.}, eq.~\eqref{eq:FluidShear} for $h<1$). These scenarios can be captured concisely by introducing an indicator function $\chi(h)$ that transitions from $\chi(h<1)=0$ to $\chi(h\equiv1)=1$, such that
\begin{equation}
	\sigma_{xz}'(z = h) = -\chi \mu_w P_s \tanh \left[ \frac{u_s(z = h)}{\varepsilon} \right] + (1 - \chi) \left[ \frac{h - 1}{2} \frac{\p P_f}{\p x} - \frac{\theta u_{s}(z = h)}{1 - {h}} \right], \label{eq:RegularisedStress2}
\end{equation}
where the first term on the right is regularised wall friction and the second is viscous shear.

In equation~\eqref{eq:RegularisedStress2}, the term proportional to $\theta$ is inversely proportional to $1-h$, which illustrates an awkward feature of the model. As discussed in \S\ref{sec:UpperLayer} above, the upper fluid layer satisfies no-slip conditions on both the stationary top wall and the moving granular layer. However, the thickness of this fluid layer vanishes as the granular layer approaches and then touches the top wall (\textit{i.e.}, as $h\to{}1$; the bridging transition). As a result, the shear stress required to drive relative motion between the granular layer and the wall --- which requires shearing the vanishing fluid layer --- can become arbitrarily large as $h\to1$, driving the system toward a no-slip condition between the granular layer and the wall. For a system of two fluid layers, this smooth approach toward no-slip between the lower layer and the wall is physically appropriate. In the present scenario, however, the lower layer should instead transition a frictional condition when it touches the upper wall; this frictional condition features an upper limit on the shear stress, allowing for sliding (\textit{i.e.}, slip) when that limit is reached. Hence, the lower layer could experience a jump in velocity and shear stress upon contacting the top wall unless this transition is regularised. In reality, the top of the granular layer is a permeable surface with grain-scale roughness. As a result, we expect fluid flow in the upper layer to transition smoothly into flow through the granular porous medium, and the shear stress at the top surface of the granular layer to transition smoothly from fluid shear to wall friction, as the irregular granular surface gradually makes contact with the top wall.

To reflect this smooth transition, we regularise the implementation of equation~(\ref{eq:RegularisedStress2}) in two ways. First, we place a non-zero minimum on the thickness of the fluid layer (\textit{i.e.}, on $1-h$) as it appears in the final term of equation~(\ref{eq:RegularisedStress2}), which reflects the fact that this construction breaks once the gap becomes comparable to the grain size. Specifically, we use the dimensionless permeability $K$ as a floor on $1-h$ in the last term in equation~(\ref{eq:RegularisedStress2}) in order to link the regularisation to the grain size ($\sqrt{K}$ would scale more directly with the grain diameter, but would also regularise more aggressively). Second, we replace the sharp indicator function $\chi(h)$ with a smoothed step function of the form $\chi = h^n$ with $n\gg 1$, thus regularising the bridging transition. The latter regularisation has the added benefit of preserving a continuous problem in $x$, avoiding the need to track the free boundary between bridged and unbridged regions. We show in Appendix~\ref{sec:regularisation}, and specifically figure~\ref{fig:regularisation}c, that our results are insensitive to the value of $n$ provided that it is large enough.

Equations~\eqref{eq:usNondim1}--\eqref{eq:usNondim3} lead to a nonlinear first-order ODE for $u_s(z)$ that also depends on $P_s$ and $\partial_x P_f$. Expressions for the latter two quantities follow from evaluating equation~\eqref{eq:sigmaXZ} at $z=0$,
\begin{equation}
    \sigma_{xz}'(z = h) - \sigma_{xz}'(z = 0) = 
h \left( \frac{\p P}{\p x} + \frac{\p h}{\p x}  \right). \label{eq:RegularisedPs}
\end{equation}
Combined with equations~\eqref{eq:RegularisedShearz=0} and \eqref{eq:RegularisedStress2}, this relationship determines $\partial_x P_f$ in the unbridged region (where $P_s = 0$) and $P_s$ in the bridged region (where $\partial_x P_f = 0$), recalling that $P = P_s + P_f$. 

Finally, the ODE for $u_s$ requires a boundary condition. The appropriate constraint follows from global conservation of mass, equation~\eqref{eq:Incompressible2}, which becomes
\begin{align}
	1 =  h \bar{u}_s - \frac{\bar{k}}{\theta} \frac{\p P_f}{\p x} + \frac{1 - h}{2} u_s(z = h), \label{eq:ConservationOfMass}
\end{align}
where $\bar{k} = K h + (1 - h)^3/12$ is the dimensionless gap-averaged permeability.

\subsection{Numerical details; boundary and initial conditions} 

To solve the model presented in \S \ref{sec:Nondimensional}, we simplify computation by shifting to a frame $X=x-t$ that moves with the piston. With this change of frame, equation~\eqref{eq:EvolutionH} becomes
\begin{align}
    \frac{\p h}{\p t} + \frac{\p}{\p X} \left[  (\bar{u}_s - 1) h \right] = 0. \label{eq:EvolutionH2}
\end{align}
We solve the problem from initial condition $h(X,0)=h_0$ and subject to boundary conditions $u_s = u_f = 0$ at $X=0$ (grains and fluid travelling with the impermeable piston) and $\p h/\p X = 0$ at $X=L_X$ (grains undisturbed in the far field), where the domain size $L_X$ is chosen to be sufficiently far ahead of the transition zone that its value has no impact on the results. We discretize the numerical domain $0 \le X \le L_X$ into $1+L_X/\Delta{X}$ nodes of equal spacing $\Delta X = 5 \times 10^{-3}$. We then solve equation~\eqref{eq:EvolutionH2} for $h$ by approximating the spatial derivative using a second-order, essentially non-oscillatory scheme and integrating in time using the forward-Euler method with a fixed time step $\Delta t = 10^{-5}$.

Evolving equation~\eqref{eq:EvolutionH2} in time requires that we compute $\bar{u}_s$ at every discrete point $X_j$ via equations~\eqref{eq:hy}--\eqref{eq:ConservationOfMass} at each time step. To do so, we discretise the vertical domain at each point $X_j$, $0 \le z \le h(X_j)$, into $N_z=250$ points of equal spacing $\Delta{z}_j=h(X_j)/(N_z-1)$. We then approximate $\p_z u_s$ and $\p_X P_s$ at each point $X_j$ as $(\p_z u_s)_j\approx(u_s(X_j,z_{i+1})-u_s(X_j,z_i))/\Delta z_j$ and $(\p_X P_s)_j \approx (P_s(X_{j+1})-P_s(X_{j}))/\Delta X$, respectively, noting that $P_s=0$ in the unbridged region. Thus, equations~\eqref{eq:hy}--\eqref{eq:ConservationOfMass} become a coupled, nonlinear system of equations in $N_z + 6$ unknowns: $P_s(X_{j})$, $(\p_X P_f)_j$, $\sigma_{xz}'(X_j,0)$, $\sigma_{xz}'(X_j,h(X_j))$, $h_{y1}(X_j)$, $h_{y2}(X_j)$, and $u_s(X_j,z_i)$ for $i = 1$ to $N_z$. We solve this system of equations via Newton's method using the solution at the previous time step as the initial guess. We then compute $\bar{u}_s(X_j)$ by averaging $u_s$ over appropriate regions in $z$.

For the regularisation parameters in equations~\eqref{eq:RegularisedShearz=0} and \eqref{eq:RegularisedStress2}, we fix $\varepsilon = 10^{-3}$ and $n = 200$; our results are insensitive to these specific values (see Appendix~\ref{sec:regularisation}).

\section{Dry model ($\mu(I)$ rheology, $\St \gg 1$) \label{sec:DryRegion}}

We first consider the `dry' limit in which the fluid is unimportant, implying that $\St \gg 1$. As discussed in \S\ref{sec:Rheology}, we therefore impose the inertial $\mu(I)$ rheology given by equations~\eqref{eq:muI} and \eqref{eq:InertialNumber} in the region where $|\p_z u_s| > 0$. The dry limit also implies that $\theta\ll{}1$, such that equation~\eqref{eq:ConservationOfMass} indicates $\p_x P_f \sim 0$. For the results that follow, we adopt the illustrative parameter values $I_0 = 0.3$, $\mu_s = 0.4$, and $\mu_2 = 0.7$, which are typical of monodisperse glass beads \citep{Jop2006,Pouliquen2006}, and we explore the roles of $\mu_w$, $\Fr$, and $h_0$. The parameters $\theta$ and $K$ drop out of the model in this limit.

\begin{figure}
	\centering
	\includegraphics[width=0.80\linewidth]{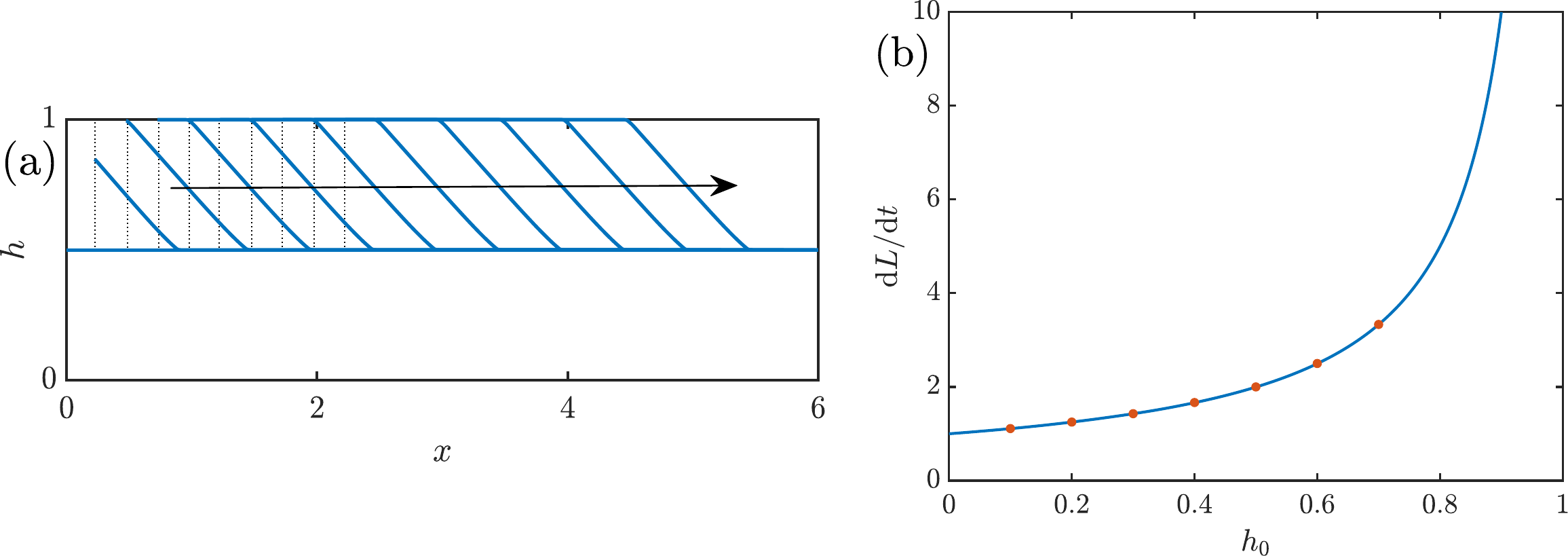} 
	\caption{(a) Height profiles of the granular layer at various times from numerical solutions to equations~\eqref{eq:EvolutionH}--\eqref{eq:ConservationOfMass} with $h_0 = 0.5$, $\Fr = 0.1$, and $\mu_w = 0.85$. The vertical, dotted black lines denote the location of the piston and the black arrow indicates direction of increasing time, with solutions being plotted at time intervals of $0.25$ for $0\leq{}t\leq{}2.25$. {(b) The translation speed of the transition region (\textit{i.e.}, the rate of change of the length of the bridged region, $\mathrm{d}L/\mathrm{d}t$) in the steady travelling-wave limit from transient numerical simulations (red dots) and from conservation of solid volume, $\dot{L} = 1/(1 - h_0)$ (blue curve).} \label{fig:exampleSolution} }
\end{figure}

To demonstrate the behaviour of the model, we present in figure~\ref{fig:exampleSolution}(a) an example simulation performed with $\Fr = 0.1$ and $\mu_w = 0.85$. Our simulation qualitatively reproduces the behaviour illustrated in figure~\ref{fig:schematic}. The grains undergo a period of initial pile-up until they bridge the gap. Thereafter, the length of the bridged region, $L$, continues to grow and the transition region evolves toward a steady travelling-wave with constant speed $\dot{L} = 1/ (1 - h_0)$ (figure~\ref{fig:exampleSolution}b), which follows from conservation of solid volume. As such, for visualisation purposes, it is convenient to introduce a travelling-wave coordinate $\tilde{x} = x - L$, such that the bridged region occupies $\tilde{x} < 0$ while the transition and undisturbed regions occupy $\tilde{x} > 0$. We consider the model to have reached a steady state once the shape of the transition region and the solid velocity within the transition region cease to vary in the travelling frame, which generally occurs quickly after the grains have bridged the gap.

We next compute the solid velocity within the steady travelling wave over a range of values of $\Fr$ for $\mu_s < \mu_w < \mu_2$ and for $\mu_s<\mu_2<\mu_w$, as shown in figure~\ref{fig:BridgedRegionDry}. The horizontal red lines indicate where the grains are sliding along the walls (\textit{i.e.}, the regions where $u_s > 0$ at $z = 0$ and $1$). Recall that $\mu(I)\to\mu_2$ for $I\gg{}I_0$, such that $\mu_2$ is the limiting value of intra-granular friction at large shear rates. We do not illustrate the case $\mu_w<\mu_s$ because it leads to the trivial velocity profile $u_s = 1$ throughout the bridged and transition regions for all $\Fr$ (figures~\ref{fig:sketchshear}c and \ref{fig:sketchshearbridge}b). Next, we investigate in detail the motion of the grains in the bridged region (\S\ref{sec:DryBridgedZone}), and in the transition (\S\ref{sec:DryTransitionZone}).

\subsection{Bridged region} \label{sec:DryBridgedZone}

\begin{figure}
	\centering
	$\mu_s < \mu_w = 0.55 < \mu_2$  \hspace{5cm} $\mu_s < \mu_2 < \mu_w = 0.85$\\
	$\Fr = 0.01$\\
	\includegraphics[width=0.45\linewidth]{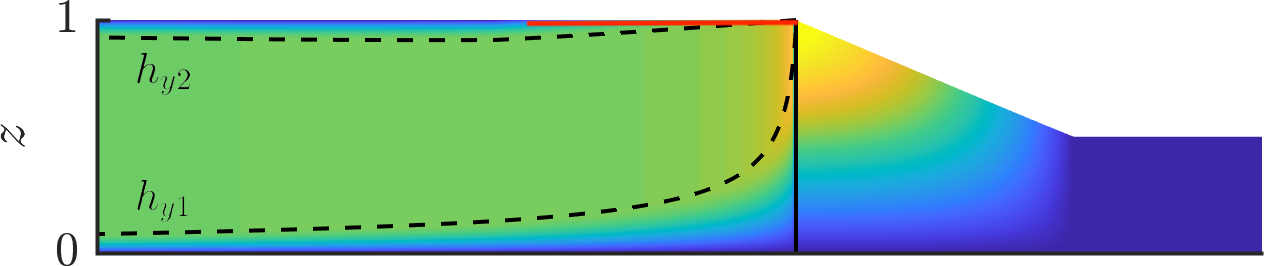} 
	\hspace{0.25cm}
	\includegraphics[width=0.42\linewidth]{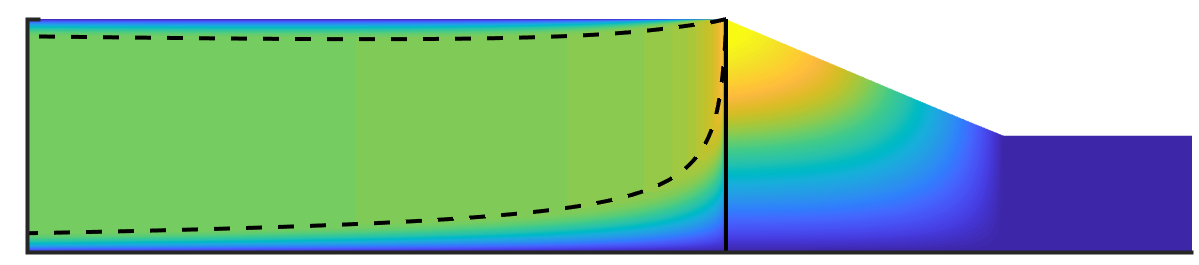}  \\
	$\Fr = 0.05$\\
	\vspace{0.25cm}
	\includegraphics[width=0.45\linewidth]{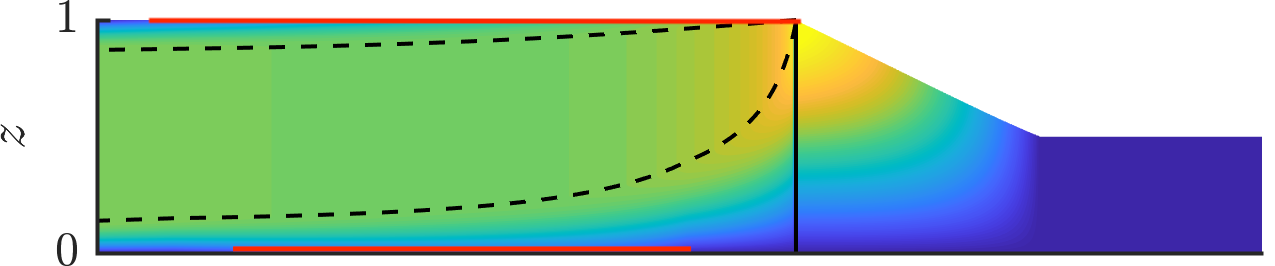} 
	\hspace{0.25cm}
	\includegraphics[width=0.42\linewidth]{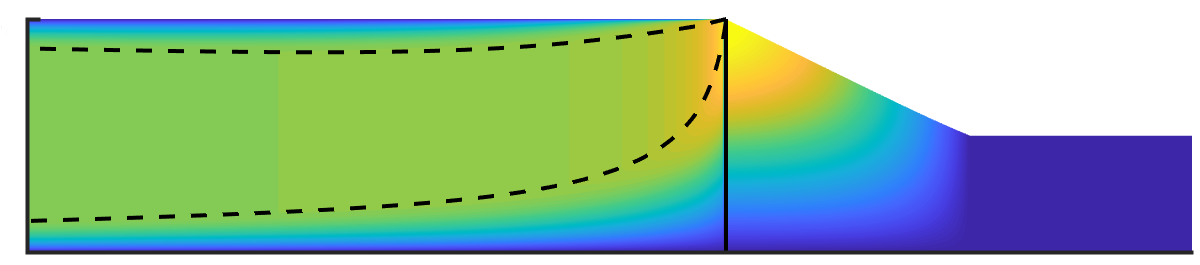} \\
	$\Fr = 0.35$\\
	\vspace{0.25cm}
	\includegraphics[width=0.45\linewidth]{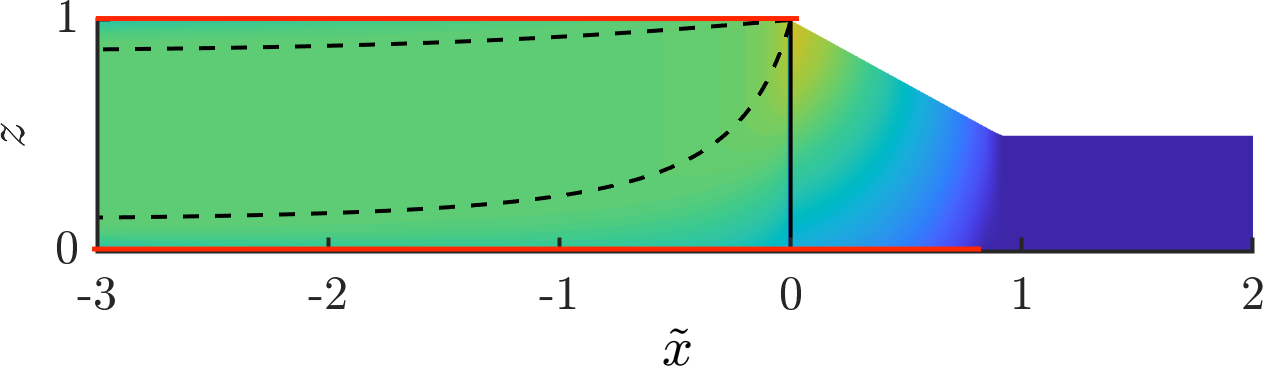} 
	\hspace{0.25cm}
	\includegraphics[width=0.42\linewidth]{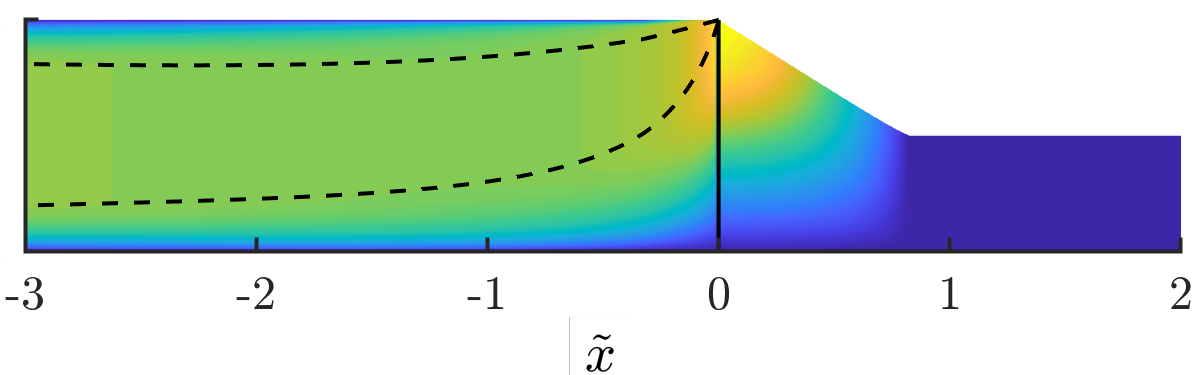}\\
	\includegraphics[width=0.40\linewidth]{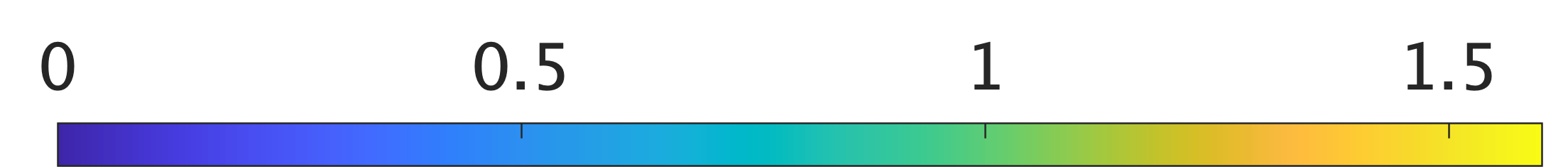}\\
    Horizontal velocity
	\caption{The velocity profile of the solid phase at steady state computed with $\mu_w = 0.55$ (left) and $0.85$ (right). The Froude number (top to bottom) is $\Fr = 0.01$, 0.05, and 0.35. The dotted black lines denote the location of $h_{y1}$ and $h_{y2}$ defined by equations \eqref{eq:hy}, recalling that $\p_z u_s = 0$ for $h_{y1} < z < h_{y2}$. The solid black vertical lines denote the end/start of the bridged/transition regions. The red lines denote the regions where the grains slide over the walls, i.e., $u_s > 0$ at $z = 0$ and 1.}
	\label{fig:BridgedRegionDry}
\end{figure}

Since $h\equiv1$ in the bridged region by definition, equation~\eqref{eq:EvolutionH} becomes unnecessary and equation~\eqref{eq:ConservationOfMass} simplifies to $\bar{u}_s = 1$. However, as discussed in \S\ref{sec:LowerLayer}, the solid velocity is nontrivial due to intra-granular friction and wall friction. As illustrated in Figure~\ref{fig:BridgedRegionDry} row 1, when $\Fr$ is sufficiently small, the grains will yield over the entire gap around the start of the bridged region ($h_{y1} = h_{y2} = 1$),  and the grains do not slide over the bottom plate. As the length of the bridged region grows, the grains stop sliding over the top plate, resulting in the velocity profile illustrated in Figure \ref{fig:sketchshearbridge}(e). When $\mu_w < \mu_2$ (column 1), as $\Fr$ increases, we find that the grains begin to slide over both the top and bottom plates. Further, for large $\Fr$, the grains appear to resist internal rearrangement such that $|\p_z u_s| \ll 1$ across the entire gap. In comparison, when $\mu_w > \mu_2$  (column 2), we find that the grains do not slide over the bottom or top plate for all $\Fr$, and the velocity profile does not appear to change significantly as $\Fr$ is varied. 

\begin{figure}
	\centering
	\includegraphics[width=0.32\linewidth]{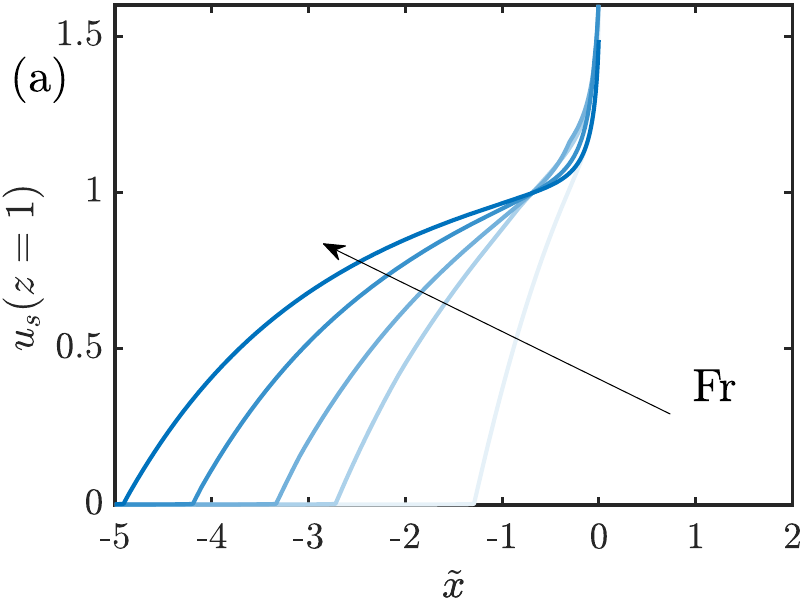}
	\includegraphics[width=0.32\linewidth]{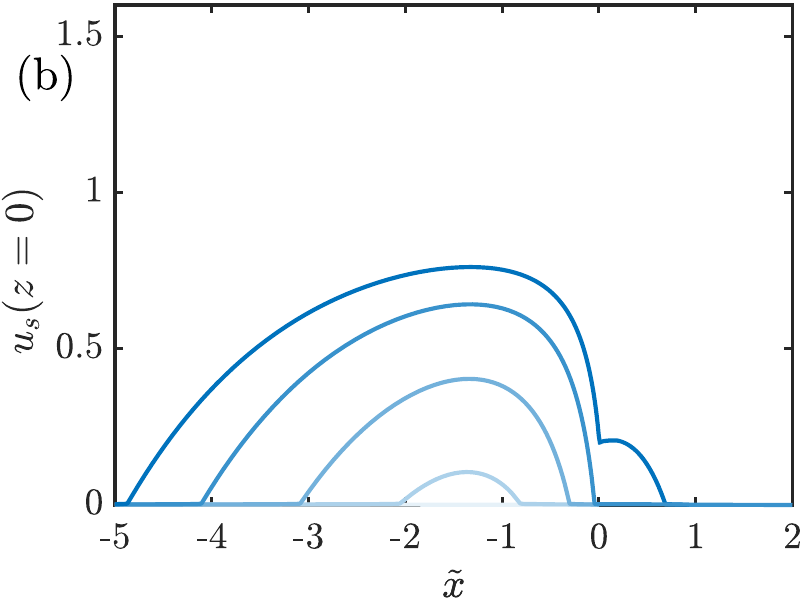} 
	\includegraphics[width=0.32\linewidth]{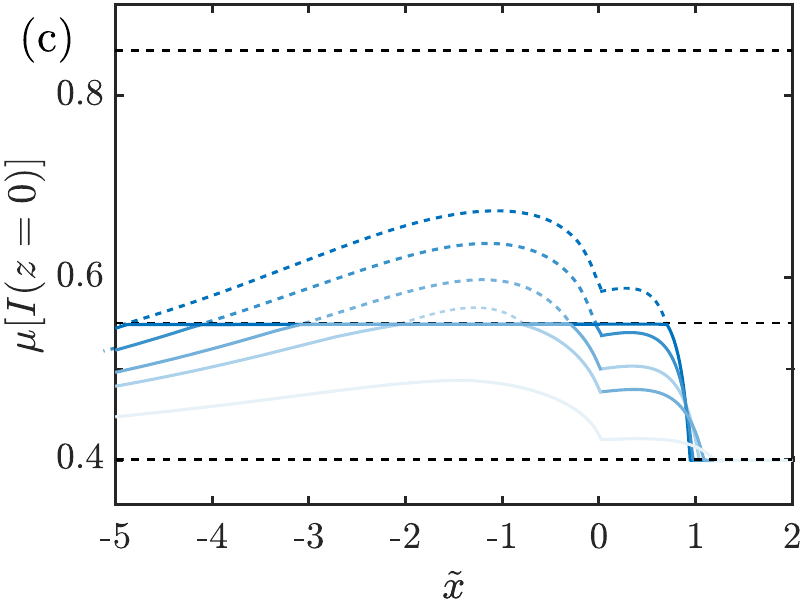}
	\caption{Velocity of the grains on the (a) top and (b) bottom plates with $\mu_w = 0.55$ and (light to dark blue) $\Fr = 0.01$, 0.04, 0.06, 0.1, and 0.15.
		Plot (c) shows the friction factor $\mu(I)$ evaluated at $z = 0$ for $\mu_w = 0.55$ (solid) and 0.85 (dashed). The dotted black lines denote $\mu_s = 0.4$, and the two values of $\mu_w$ being considered.}
	\label{fig:wallvelocitybottommuw0}
\end{figure}

To better understand the influence of the Froude number and the wall friction parameter, we compute the velocity profile of the grains at $z = 1$ and $0$ over a range of values of $\Fr$ with $\mu_s < \mu_w = 0.55 < \mu_2$, shown in Figures~\ref{fig:wallvelocitybottommuw0}(a) and (b). The velocity of the grains in contact with the top plate [Fig~\ref{fig:wallvelocitybottommuw0}(a)] is fastest at the start of the bridged region and monotonically decreases to $0$ as the length of the bridged region grows, after which the grains are stationary. Further, the length of the region where grains slide over the top plate increases with $\Fr$. For the grains in contact with the bottom plate [Fig~\ref{fig:wallvelocitybottommuw0}(b)], $\Fr$ must be sufficiently large or the grains will be stationary over the entire bridged region. When $\Fr$ becomes sufficiently large the grains will begin to slide, which corresponds to the point where $\mu(I) = \mu_w$ shown in Figure \ref{fig:wallvelocitybottommuw0}(c). However, unlike grains in contact with the top plate, grains will be stationary at the start of the bridged zone, resulting in a region where grains slide sandwiched between regions where the grains are stationary. As $\Fr$ increases further, this region where $u_s(z = 0) > 0$ will `spread' and the grains at the start of the bridged zone will begin to move (red curves). In comparison, when $\mu_w = 0.85$, grains do not slide over either plate for all $\Fr$.  Figure~\ref{fig:wallvelocitybottommuw0}(c) indicates that while $\mu(I)$ (denoted by dashed lines for $\mu_w = 0.85$) increases with $\Fr$ as it does for $\mu_w = 0.55$, it will always be smaller than $\mu_w$, in this case, because it is bounded above by $\mu_2 = 0.7 < \mu_w$.

\begin{figure}
	\centering
	\includegraphics[width=0.30\linewidth]{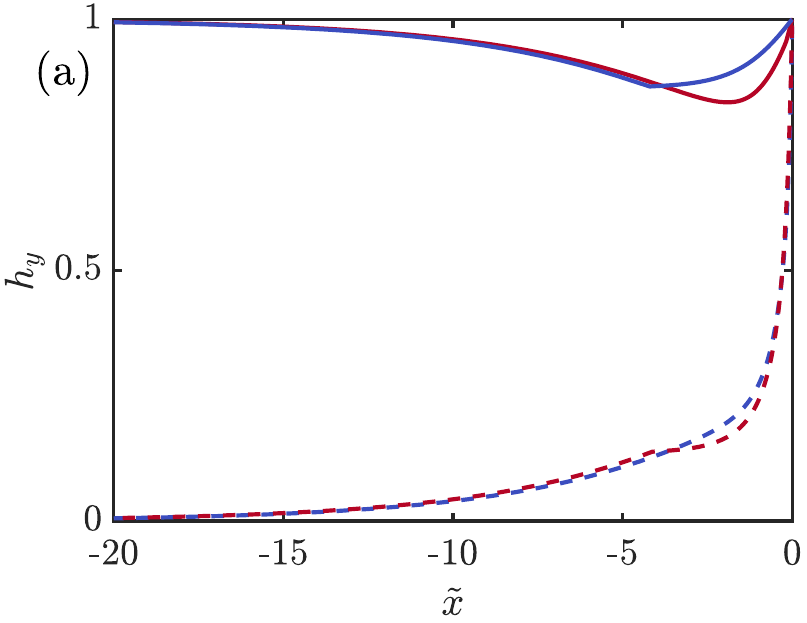}
	\includegraphics[width=0.30\linewidth]{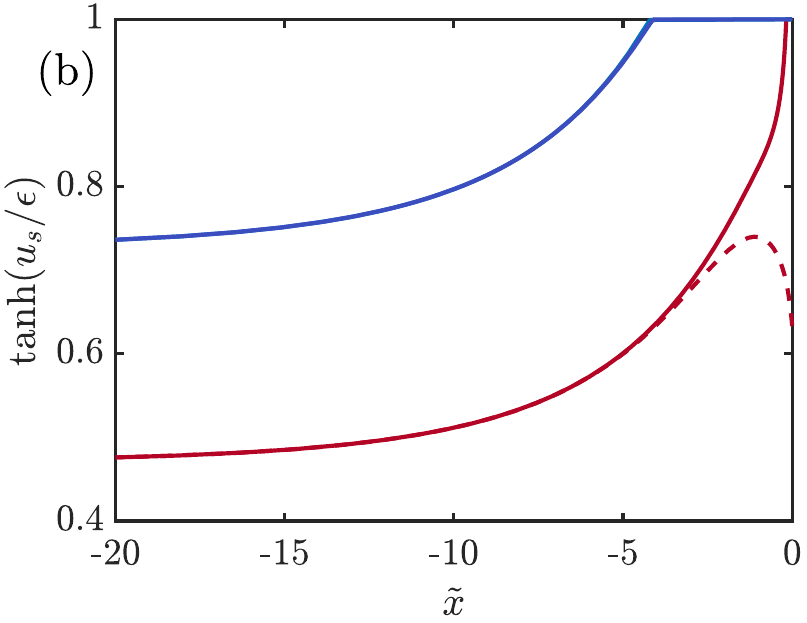}
	\includegraphics[width=0.30\linewidth]{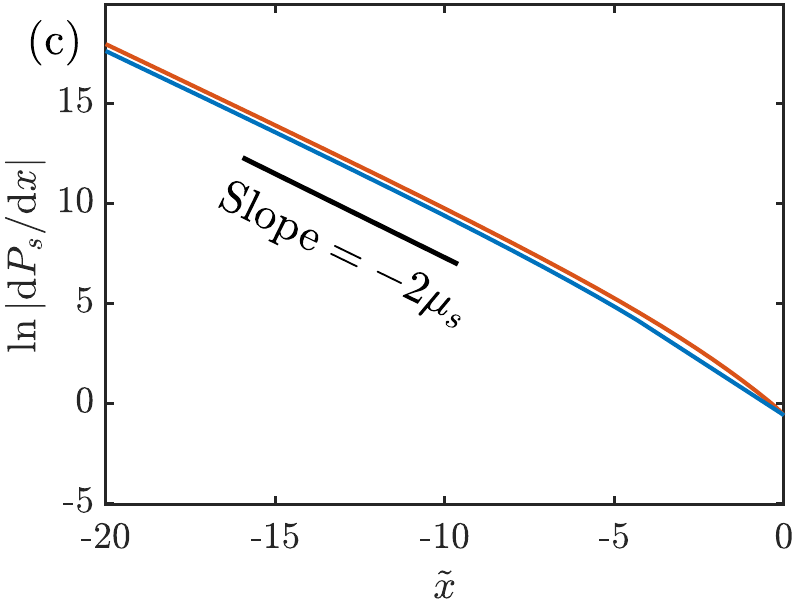}
	\caption{(a) The location of the yield layers $h_{y1}$ (dashed curve) and $h_{y2}$ (solid curve) determined by equations \eqref{eq:hy}. (b) The hyperbolic tangent of the solid velocity at $z = 1$ (solid curve) and $z = 0$ (dashed line), which are used to compute the regularised shear stress according to equations \eqref{eq:RegularisedShearz=0} and \eqref{eq:RegularisedStress2}, recalling that $\varepsilon = 10^{-3}$. (c) Natural log of $|\p_x P_s|$ computed from equation \eqref{eq:RegularisedPs}. Solutions are computed with $\Fr = 0.1$ and $\mu_w = 0.55$ (blue) and $0.85$ (red).}
	\label{fig:tanh}
\end{figure}

As the length of the bridged region grows and $P_s$ becomes sufficiently large, for all $\Fr$ and $\mu_w > \mu_s$, the grains will not be able to slide over both the top and bottom plates. Further, as indicated by Figure \ref{fig:tanh}(a), $h_{y1}$ and $h_{y2}$ evolve from 0 to 1, respectively, toward 1. As such, the velocity profile of the grains will tend to $u_s \approx 1$, except for the increasingly thin regions around $z = 0$ and 1. Such a profile is not observed in experiments, discussed further in section \ref{sec:Experiments}, and we anticipate that our continuum approximation will break down when $h_{y1}$ and $1 - h_{y2}$ become sufficiently small. When the velocity of the grains in contact with the walls are of order $\varepsilon$, Figure \ref{fig:tanh}(b) shows that $\tanh(u_s/\varepsilon)$ at $z = 0$ and 1 will tend to $\mu_s / \mu_w$, which still satisfies $u_s \ll 1$. Thus from equation \eqref{eq:RegularisedPs}, it follows that $\p_x P_s \sim -2\mu_s P_s$, indicating that pressure builds up exponentially through the bridged region, as illustrated in Figure~\ref{fig:tanh}(c). By comparison, when $\mu_w < \mu_s$ where we have the trivial velocity profile $u_s = 1$, $\tanh(u_s / \varepsilon) \sim 1$ and $\p_x P_s \sim -2\mu_w P_s$ resulting in the same exponential behaviour, but not controlled by the wall friction factor. This exponential relationship between the length of the bridged region and the normal stress at the piston is consistent with previous DEM simulations \citep{Marks2015}.

\subsection{Transition zone} \label{sec:DryTransitionZone}

We now turn our attention to the behaviour of the transition zone, which, as illustrated in Figure~\ref{fig:schematic}(c), is the wedged-shape region with $h < 1$ and $P_s = 0$. In this region, given that the impact of the fluid on the grains is negligible, it follows that $\sigma_{xz}(z = h) \sim 0$, and the (non-dimensional) shear stress, equation \eqref{eq:sigmaXZ}, reduces to
\begin{align}
	\sigma_{xz}' = (z - h) \frac{\p h}{\p x}.
\end{align}
As illustrated in Fig.~\ref{fig:sketchshear}, when $\sigma_{xz}(z = h) = 0$ the entire solid layer will be plugged ($h_{y1} = 0$ and $h_{y2} = h$) if $\mu_w < \mu_s$. If $\mu_w > \mu_s$, the solid layer will yield ($h_{y1} = h_{y2} = h$) if $|\p_x h| > \mu_s$ or be plugged otherwise. As such, we can reduce equation \eqref{eq:PiecewiseVelocity} to 
\begin{align}
	u_s &= \begin{cases}
			u_{s}(z = 0)  & \textrm{if } 0 \le z < h_y \\
			u_{sy}(x,z,t)   & \textrm{if } h_y \le z < \le h
		\end{cases} ,
\end{align}
where
\begin{align}
	h_y &= \begin{cases}
			0  & \textrm{if } |\p_x h| > \mu_s \\
			h   & \textrm{otherwise}
		\end{cases}.
\end{align}
As such, we arrive at the so-called Bagnold profile \citep{Andreotti2013} in the yielded region
\begin{align}
	u_{sy} = u_s(z = 0) + \frac{2 I_0}{3 \Fr} \frac{\p_x h + \mu_s}{\p_x h + \mu_2}  \left[ (h - z)^{3/2} - (h - h_y)^{3/2} \right] \label{eq:Bagnold},
\end{align}
where
\begin{align}
	0 = \frac{\p h}{\p x} + \mu_w \tanh \left( \frac{u_s(z = 0) }{\varepsilon} \right) \label{eq:DryModel3}.
\end{align}
Further, by substituting the inertial number, $I = \Fr \cdot \p_z u_s / \sqrt{h-z}$, into equation \eqref{eq:muI}, we arrive at
\begin{align}
	I = -\frac{I_0 (\p_x h  + \mu_s)}{\p_x h  + \mu_2}, \label{eq:SimplifiedInertial}
\end{align}
which combined with equation \eqref{eq:muI} reduces to $\mu(I) = -\p_x h$.

Recall that representative solutions were shown in Figure \ref{fig:BridgedRegionDry} with $\mu_w < \mu_2$ in column 1 and $\mu_2 < \mu_w$ in column 2. We find that as $\Fr \to 0$ (row 1), the height of the granular pile tends to a line of slope $-\mu_s$. Further, the grains on the bottom plate are stationary such that the velocity profile takes the form illustrated in Figure \ref{fig:sketchshear}(b). When $\mu_w < \mu_2$ (column 1), as $\Fr$ increases, the length of the transition zone decreases and the grains begin to slide along the bottom plate. For sufficiently large $\Fr$ (row 3), the grains appear to resist internal rearrangement such that $|\p_z u_s| \ll 1$. When $\mu_2 < \mu_w$ (column 2), again the slope of the transition zone decreases with increasing $\Fr$. However, unlike in column 1, the grains remain stationary on the bottom plate for all $\Fr$. Analogous behaviour was observed in the bridged region discussed previously. For the case where $\mu_w < \mu_s$ (not shown) the grains do not yield; as such, solutions are independent of $\Fr$ and the granular pile takes the shape of a wedge with $\p_x h = -\mu_w$.

\begin{figure}
	\centering
	\includegraphics[width=0.35\linewidth]{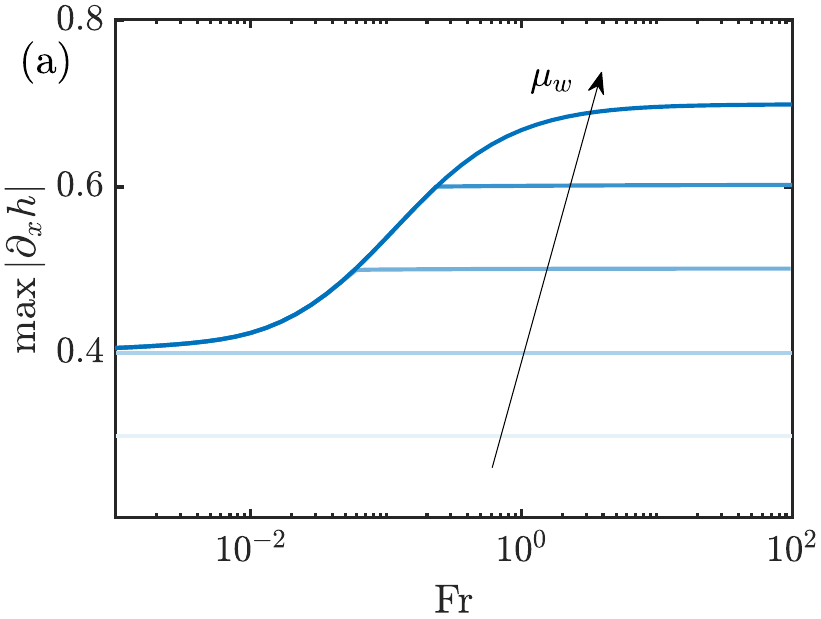}
	\includegraphics[width=0.35\linewidth]{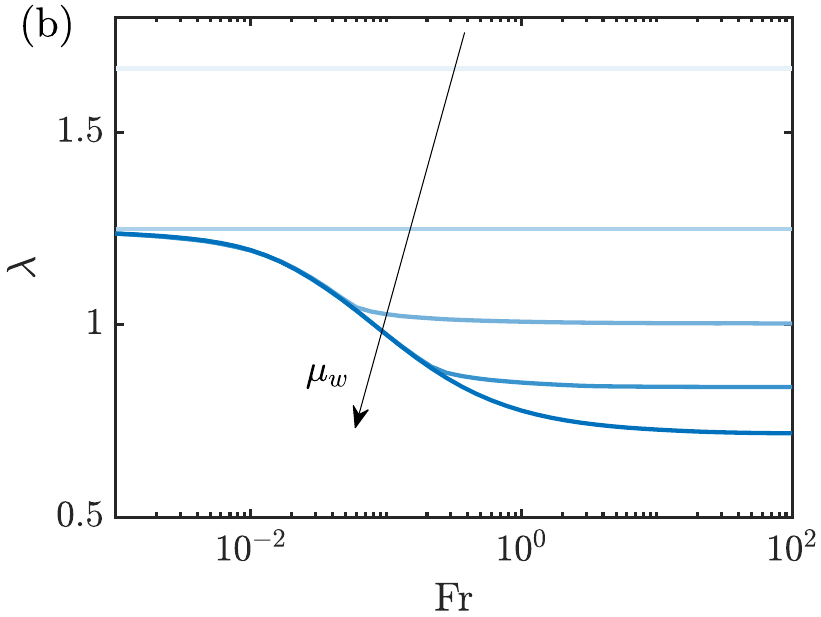}
	\caption{(a) The maximum magnitude of the slope of $h$, recalling that $I = -\p_x h$,  computed with $h_0 = 0.5$, $\mu_s = 0.4$, and $\mu_w = 0.3$, 0.4, $\ldots$, $0.7$. (b) The corresponding length $\lambda$ of the transition region.}
	\label{fig:lambdaDry}
\end{figure}

We compute the maximum magnitude of the slope of $h$ as a function of $\Fr$ over a range of values of $\mu_w$ [Figure \ref{fig:lambdaDry}(a)], as well as the corresponding length $\lambda$ of the transition zone [Figure \ref{fig:lambdaDry}(b)], where we define the latter as the distance between the last bridged point and the first undisturbed point. We determine these values by integrating our full time-dependent model for a sufficiently long time such that the shape and velocity of the transition zone become constant. We find that for all $\mu_w > \mu_s$, $\max|\p_x h|$ --- and in turn $\max|\mu(I)|$ --- will tend to $\mu_s$ as $\Fr \to 0$. This behaviour can be explained by taking $\Fr \to 0$ with $u_{s1} \sim \mathcal{O}(1)$ in equation \eqref{eq:Bagnold}, which implies that $\p_x h \sim -\mu_s$. When $\mu_s < \mu_w < \mu_2$, $\max|\p_x h|$ (and $\max|\mu(I)|$) will monotonically increase with $\Fr$ from $\mu_s$ and plateau at $\mu_w$. Once $\max|\p_x h|$ has plateaued, the shear stress is sufficiently large on the base such that the grains will begin sliding, $u_s(z = 0) > 0$. Given that the inertial number, $I \sim \Fr \cdot \p_z u_s$, does not increase above $\mu_w$ once grains begin sliding along the base, it follows that $\p_z u_s \to 0$ as $\Fr$ becomes large, resulting in the plug-like flow observed in row 3 of Figure \ref{fig:BridgedRegionDry}. Further, taking the limit that $\Fr$ is large, equations \eqref{eq:Bagnold} and \eqref{eq:DryModel3} imply that $u_s \sim u_s(z = 0)$ and $\p_x h \sim -\mu_w$. When $\mu_w > \mu_2$, we find that $\max|\p_x h|$ will monotonically increase from $\mu_s$ to $\mu_2$ with increasing $\Fr$, while $\lambda$ will decrease from $\lambda \to 1/(2\mu_s)$ to $1/(2\mu_2)$. As $\max|\p_x h| < \mu_w$ for all $\Fr$, the shear stress at the base is never sufficiently large for the grains to slide over the bottom plate and remain stationary. As such, solutions are independent of $\mu_w$ for all $\mu_w > \mu_2$. In the case where $\mu_w < \mu_s$, as discussed above, we find that $\max|\p_x h| = \mu_w$ and $\lambda = 1/(2\mu_w$) for all $\Fr$. Note that it is no longer reasonable to neglect the macroscopic inertia of fluid--solid mixture once $\Fr \gtrsim O(1)$, as discussed in \S\ref{sec:Nondimensional}, so inertial effects should play a role for the solutions with the largest $\Fr$ here.

\begin{figure}
	\centering
	\includegraphics[width=0.35\linewidth]{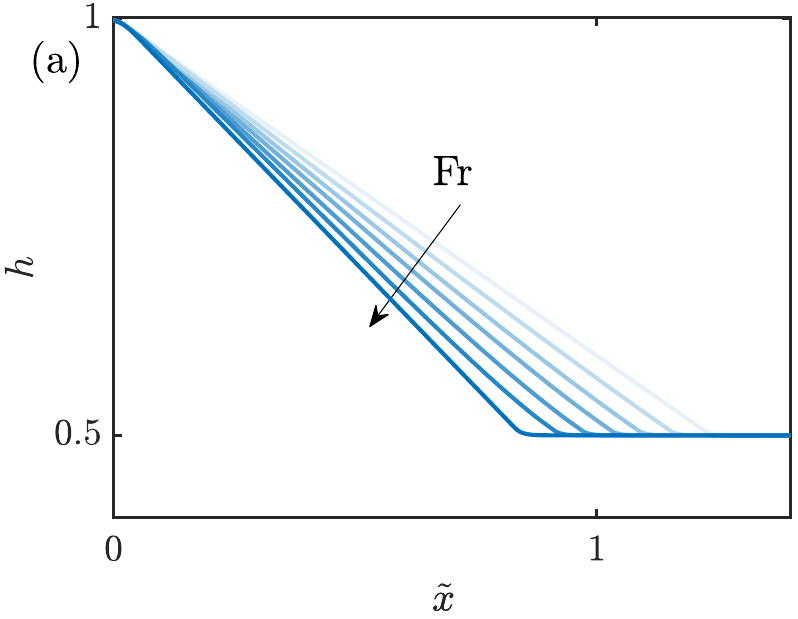}
	\includegraphics[width=0.35\linewidth]{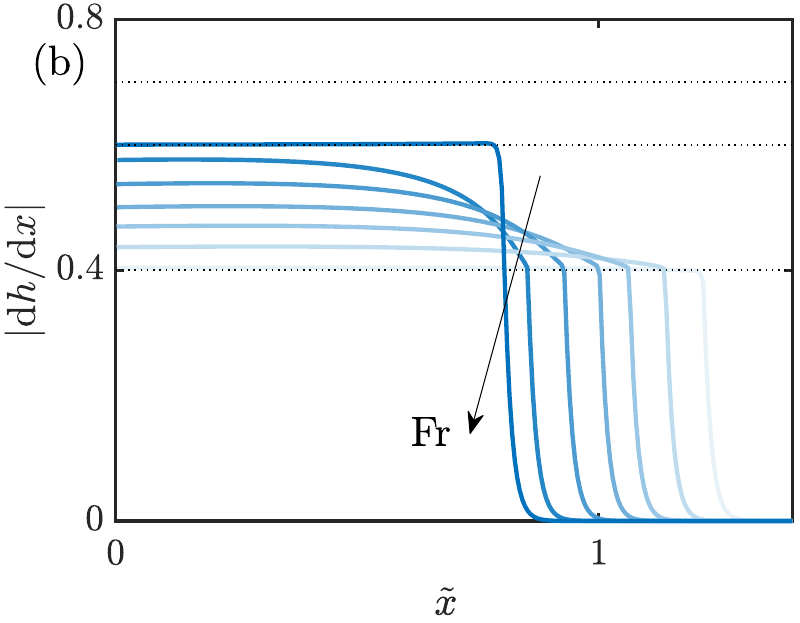}
	\caption{(a) The height of the granular pile for $h_0 = 0.5$, $\mu_w = 0.6$, and $\Fr = $ $0.001$, $0.017$, $0.036$, $0.06$, $0.1$, $0.17$, and $1$. (b) Corresponding magnitude of the slope of $h$. Dotted black lines in (b) denote values of $\mu_s$, $\mu_w$, and $\mu_2$.}
	\label{fig:hShape}
\end{figure}

Although $h$ will tend to a straight line in the limits such that $\Fr$ becomes sufficiently small or large, the shape of the granular pile is not trivial for intermediate values of $\Fr$. To better understand how the shape of the transition zone varies, we solve for $h$ over a range of values of $\Fr$ [Figure \ref{fig:hShape}(a)], and extract the corresponding slope [Figure \ref{fig:hShape}(b)]. For small $\Fr$, the magnitude of the slope of $h$ tends to $\mu_s$ as expected. As $\Fr$ increases, $h$ becomes slightly more non-linear and $|\p_x h|$ non-uniformly increases, where the slope is steeper at the start of the transition zone and decreases to 0 by the end. The magnitude of the slope increases with $\Fr$ until it reaches $\mu_w$, which corresponds to the point at which the grains will slide along the bottom plate. We note that the shape of the transition zone is in good qualitative agreement with previous results. Specifically, \citet{Marks2015}, who performed DEM simulations of bulldozing in a confined geometry, showed that the shape of the transition zone is approximately linear with a gradient that steepens as the equivalent of the Froude number increases. Similarly, \citet{Sauret2014}, who investigated a continuum model of dry bulldozing in an unconfined geometry, showed that profiles are primarily linear in shape but can become slightly `curvy', particularly around the leading edge, with increasing $\Fr$.

\begin{figure}
	\centering	
	\includegraphics[width=0.35\linewidth]{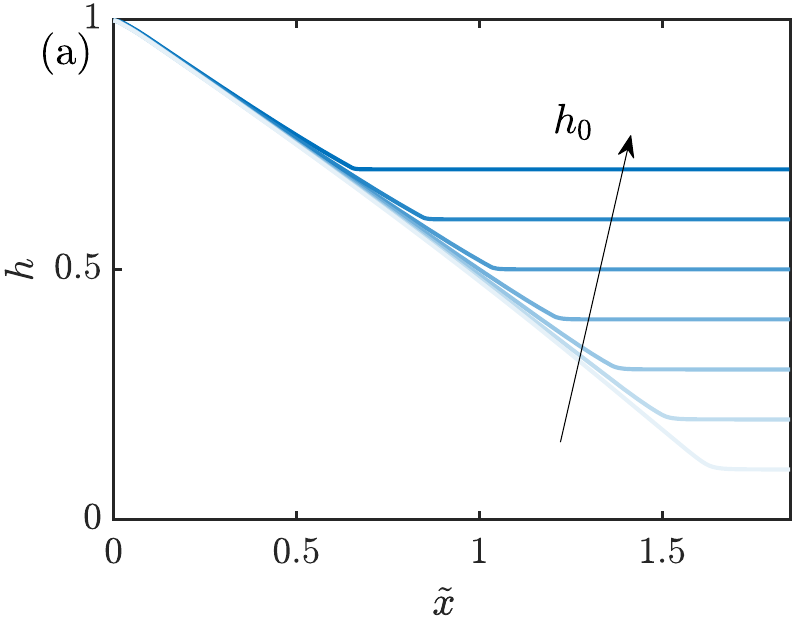}
	\includegraphics[width=0.35\linewidth]{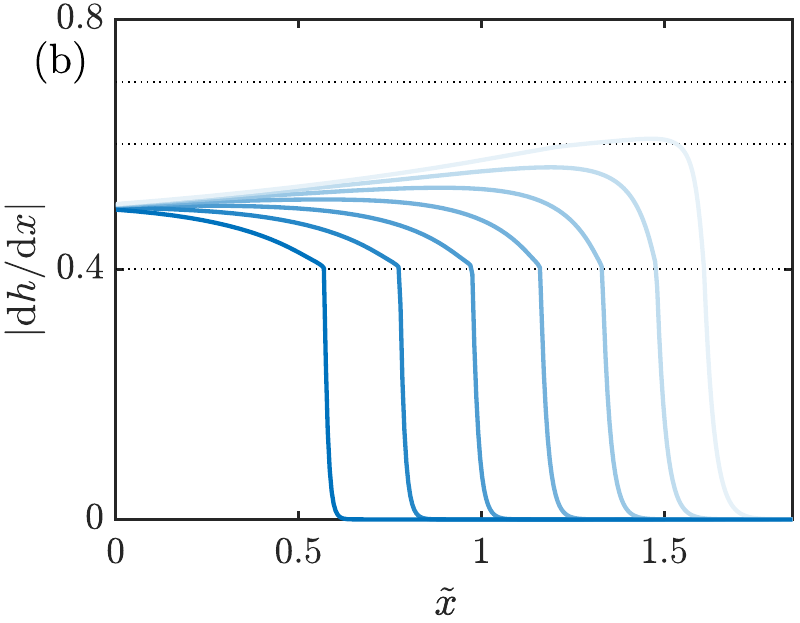}
	\caption{(a) The height of the granular pile for $\Fr = 0.06$, $\mu_s = 0.4$, $\mu_w = 0.6$, and $h_0 = $ 0.1, 0.2, 0.3, 0.4, 0.5, 0.6, and 0.7. (b) Corresponding magnitude of the slope of $h$.  Dashed black lines denote the values of $\mu_s$, $\mu_w$, and $\mu_2$.}
	\label{fig:hShape_h0}
\end{figure}

We repeat this process for a fixed $\Fr$ across a range of values of $h_0$ [Figures \ref{fig:hShape_h0}]. For larger values of $h_0$ (darker blue curves), $|\p_x h|$ monotonically decreases over the transition zone. For smaller values of $h_0$ (blue curves), $|\p_x h|$ is non-monotonic such that it initially increases and reaches a peak and then decreases to 0. For sufficiently small $h_0$, $\max|\p_x h|$ rises above $\mu_w$, suggesting that there will be a region where the grains slide along the base ($|\sigma_{xz}'| > \mu_w p_s$ at $z = 0$) sandwiched between regions where the grains are stationary, $u_s(z = 0) = 0$.

\section{Viscous model ($\mu(I_v)$ rheology)} \label{sec:ViscousModel}

\begin{figure}
 	\centering
 	$\theta = 5 \times 10^{-6}$ \hspace{5.5cm}	$\theta = 5 \times 10^{-5}$ \\
 	\includegraphics[width=0.47\linewidth]{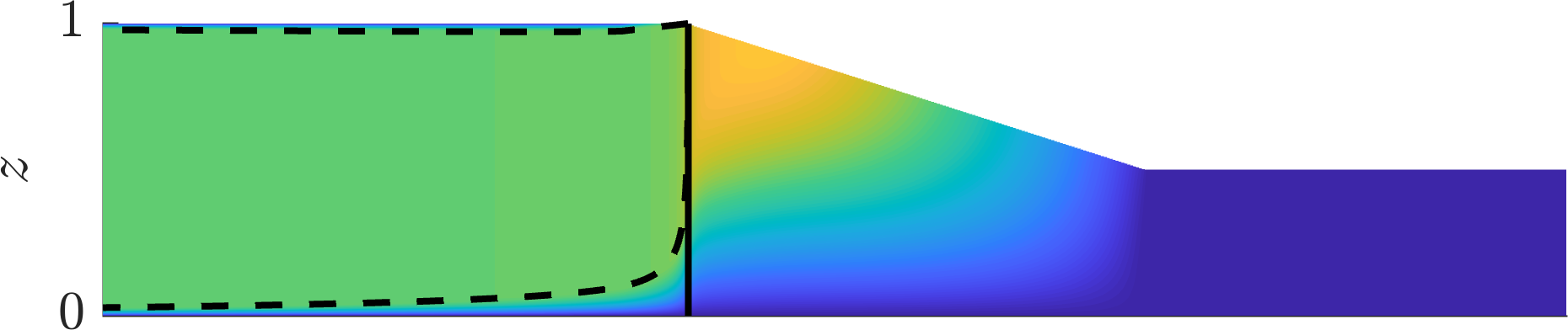} 
 	\hspace{0.5cm}
 	\includegraphics[width=0.45\linewidth]{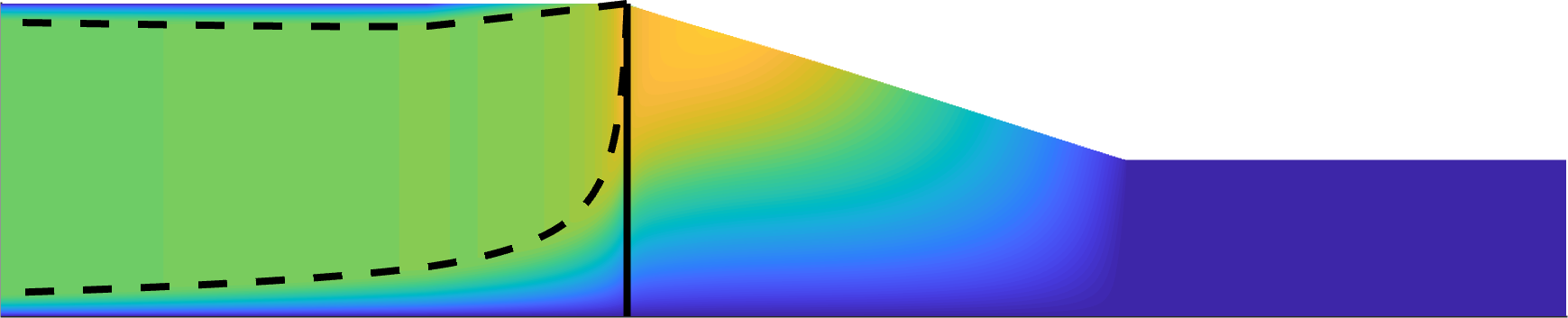} 
 	\\
 	\vspace{0.25cm}
 	$\theta = 2.5 \times 10^{-4}$  \hspace{5.5cm}	$\theta = 5 \times 10^{-4}$ \\
 	\includegraphics[width=0.47\linewidth]{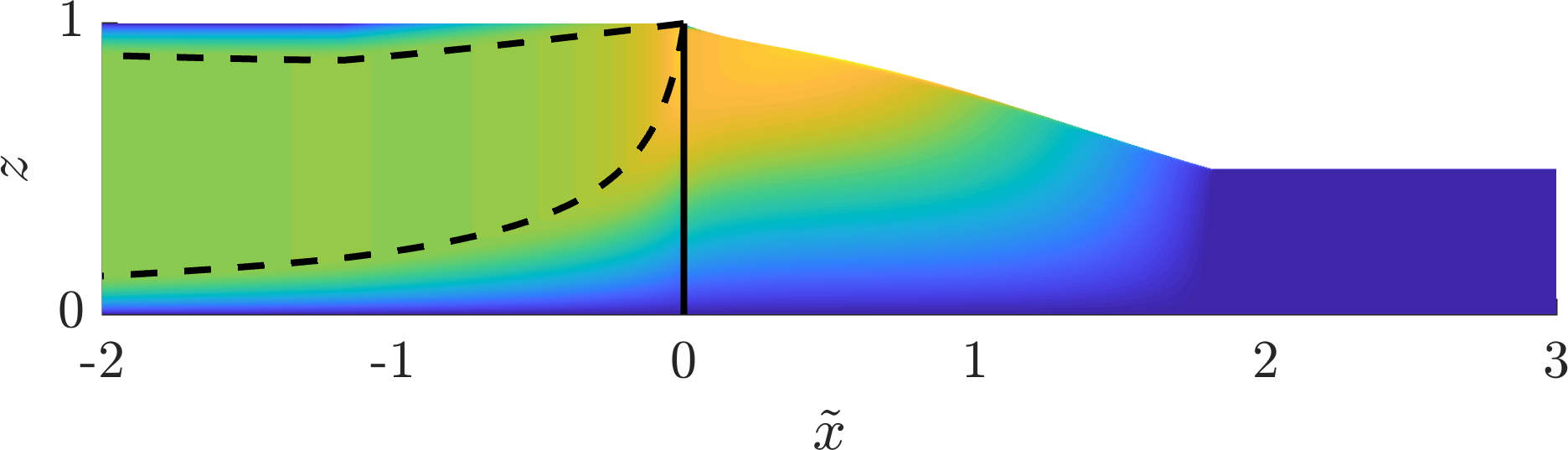} 
 	\hspace{0.5cm}
 	\includegraphics[width=0.45\linewidth]{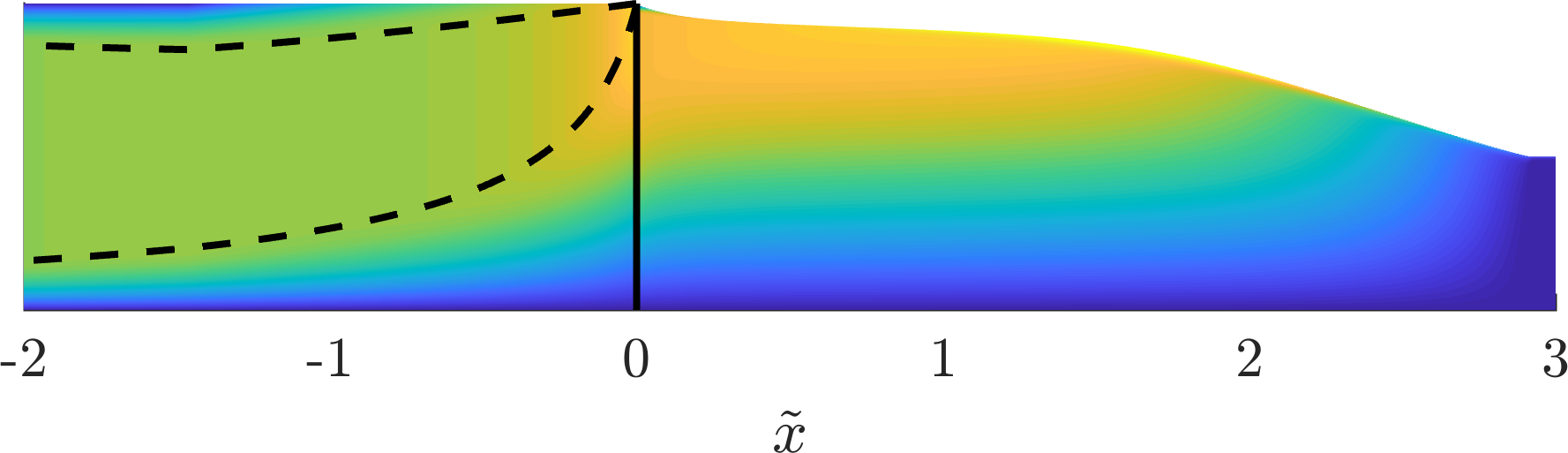}   \\
 	\includegraphics[width=0.40\linewidth]{colourbar} \\
    Horizontal velocity
 	\caption{The horizontal velocity profile of the solid phase at steady state computed with $\mu_w = 0.55$ for different values of $\theta$. The dotted black lines denote the location of $h_{y1}$ and $h_{y2}$ defined by equations \eqref{eq:hy}, recalling that $\p_z u_s = 0$ for $h_{y1} < z < h_{y2}$. }
 	\label{fig:BridgedRegionViscous}
 \end{figure}

We turn our attention to the influence of the ambient fluid on the grains. In this case, the Stokes number, recalling $\St \sim 1/\eta_f$, can be small, and as such we impose $\mu(I_v)$ rheology in the region where the grains have yielded, as discussed in \S\ref{sec:Rheology}.
As illustrative values for the following results, we take $K = 10^{-4}$, $I_0 = 0.005$, $\mu_s = 0.32$ and $\mu_2 = 0.7$, which were measured by \citet{Boyer2011} in the case of plastic beads, and we fix $h_0 = 0.5$. Recall that the model becomes independent of $\Fr$ in this limit, provided that $\mathcal{F}^2 = (\Fr\, b/d)^2 \ll 1$ to ensure that macroscopic inertia plays no role. Following Figure~\ref{fig:BridgedRegionDry}, we compute the steady state velocity profiles over a range of values of $\theta$, shown in Figure~\ref{fig:BridgedRegionViscous} (recall, $\theta$ provides a measure of the relative importance of fluid shear forces, and increases with both the piston speed and the fluid viscosity; see~(\ref{eq:defs1})). In the bridged region to the left of the vertical black line, equations \eqref{eq:ConservationOfMass} and \eqref{eq:Darcy} lead to $1 = \phi_s^* \bar{u}_s + (1 - \phi_s^*) \bar{u}_f$, which implies that $\bar{u}_s = \bar{u}_f$ and $\p_x P_f = 0$. As such, the ambient fluid does not have a direct impact on the behaviour of the grains in the bridged region. While imposing $\mu(I_v)$ rheology results in a different velocity profile to $\mu(I)$, we observe the same general behaviours with both models. That is, we find a non-monotonic profile with a plugged region sandwiched between two regions where the grains slide over one another.  However, we find the behaviour of the grains to be noticeably different in the transition region. In particular, Figure \ref{fig:BridgedRegionViscous} shows that the length of the transition region varies non-monotonically with increasing $\theta$, initially decreasing (although this is hard to see by eye) before sharply increasing.  In fact, we are unable to generate steady velocity profiles at all above a certain value of $\theta$, as discussed below.

A useful metric for quantifying the role of the fluid is the (non-dimensional) horizontal force exerted by the fluid on the grains at the interface between the upper and lower layers,
\begin{align}
	F_f = \left| \frac{\p P_f}{\p x} \right| + \frac{\sigma_{xz}'(z = h)}{h}, \label{eq:FluidForce}
\end{align}
where $\max|\mu(I_v)| \approx \max| F_f - \p_x h|$. As a point of comparison, we also computed solutions in which the overlying fluid was ignored, and thus $F_f=0$, by solving equations~\eqref{eq:EvolutionH}--\eqref{eq:ConservationOfMass} with $\p_x P_f = \sigma_{xz}'(z = h) = 0$. 

\begin{figure}
	\centering
	\includegraphics[width=0.32\linewidth]{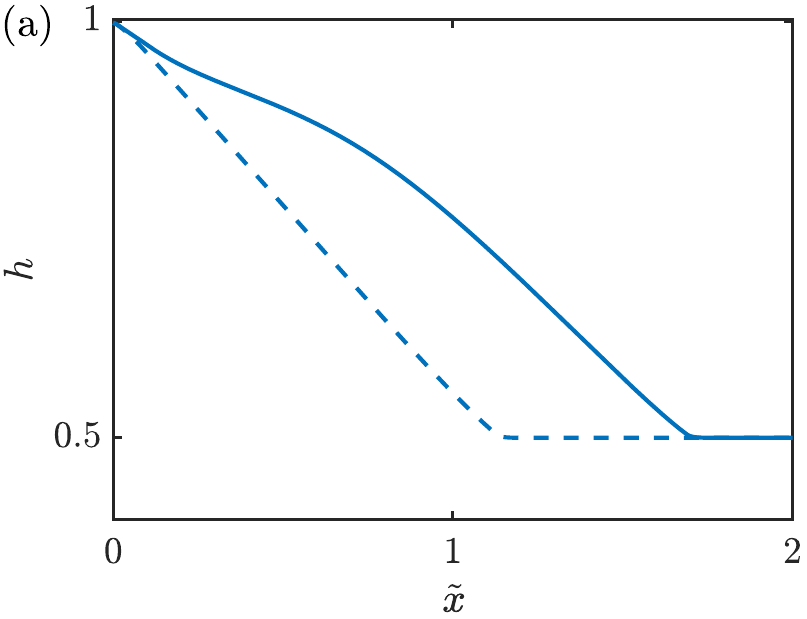}
	\includegraphics[width=0.32\linewidth]{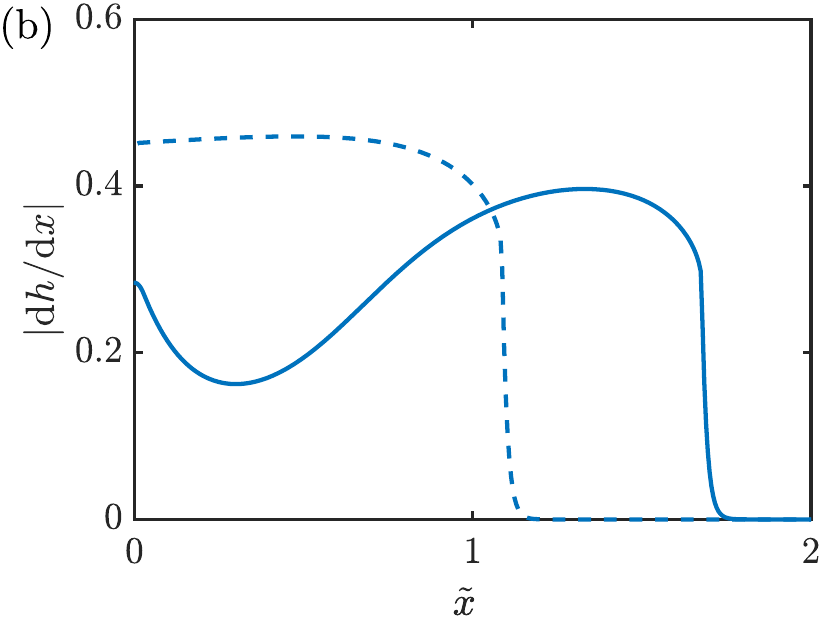}
	\includegraphics[width=0.32\linewidth]{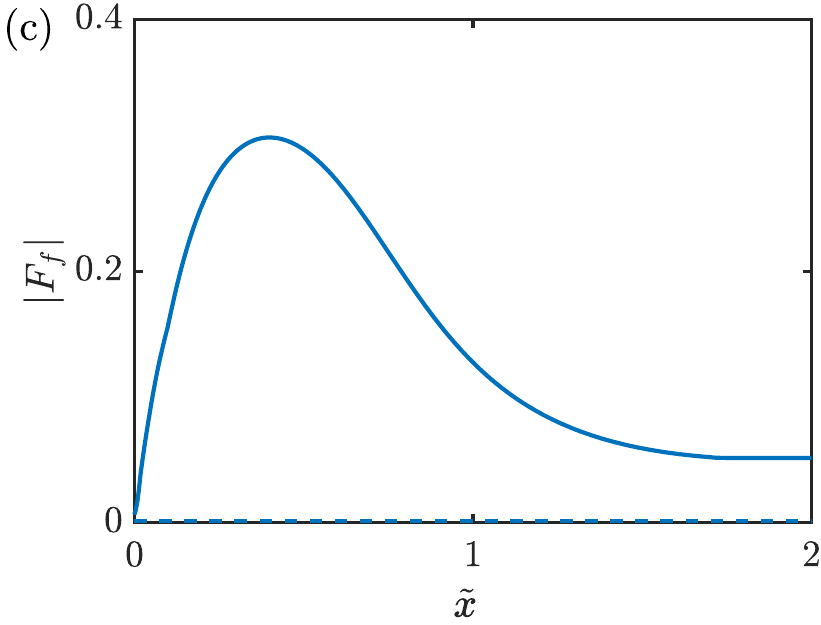}
	\caption{(a) The height of the granular pile computed with $\theta = 3.6 \times 10^{-4}$ and $\mu_w = 0.6$. Plots (b) and (c) are the corresponding magnitude of $\textrm{d}h/\textrm{d}x$ and fluid force, equation \eqref{eq:FluidForce}. 
	The dashed line denotes the case where $F_f$ is ignored by setting $\p_x P = \sigma'_{xz}(z = h) = 0$.  Note that this is distinct from the dry rheology because viscosity is still important for the internal rearrangement of the grains.}
	\label{fig:exampleSolution1}
\end{figure}

Figure~\ref{fig:exampleSolution1} shows details of the steady travelling wave from an example solution for moderate $\theta$. The length of the transition zone (figure \ref{fig:exampleSolution1}a) is larger when the fluid is included (solid lines) compared to when the fluid is ignored (dashed lines). There is a dip in the associated surface slope near the back of the transition zone (figure~\ref{fig:exampleSolution1}b), which corresponds to a peak in force from the overlying fluid $F_f$ (figure~\ref{fig:exampleSolution1}c).  Physically, increasing viscous forces in the fluid phase makes it harder for it to be displaced by the solid as the grains pile up. This effect is exacerbated as $\theta$ increases, which causes the transition zone to be elongated. 

\begin{figure}
	\centering
	\includegraphics[width=0.39\linewidth]{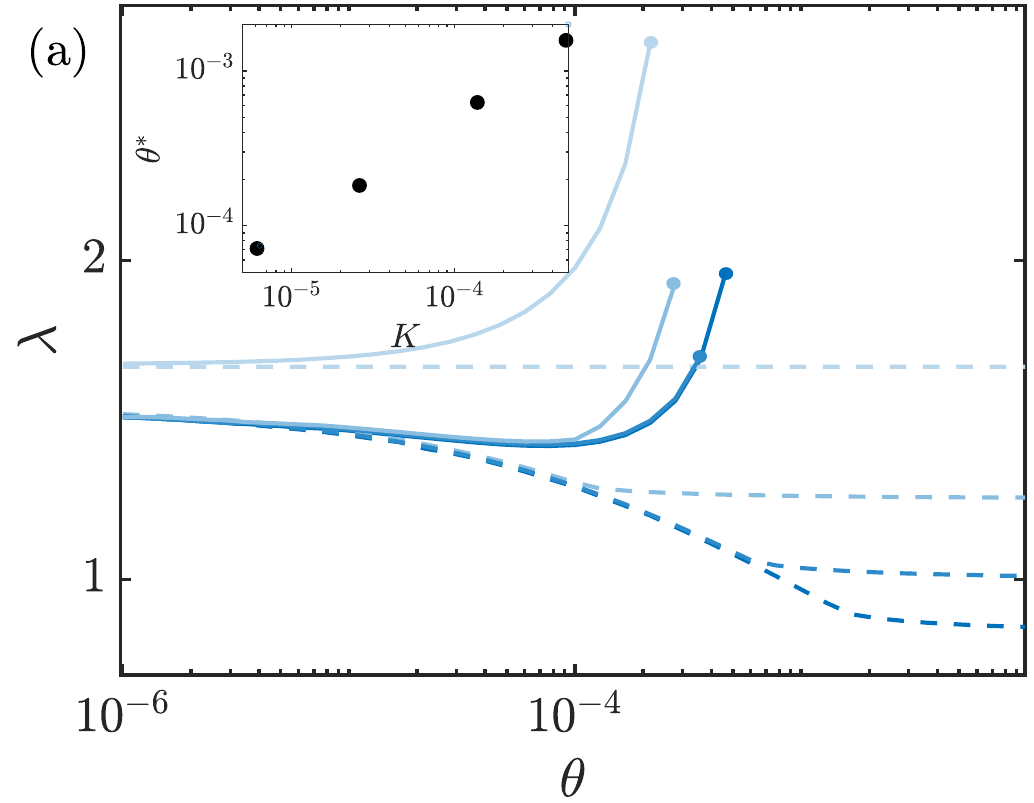}
	\includegraphics[width=0.4\linewidth]{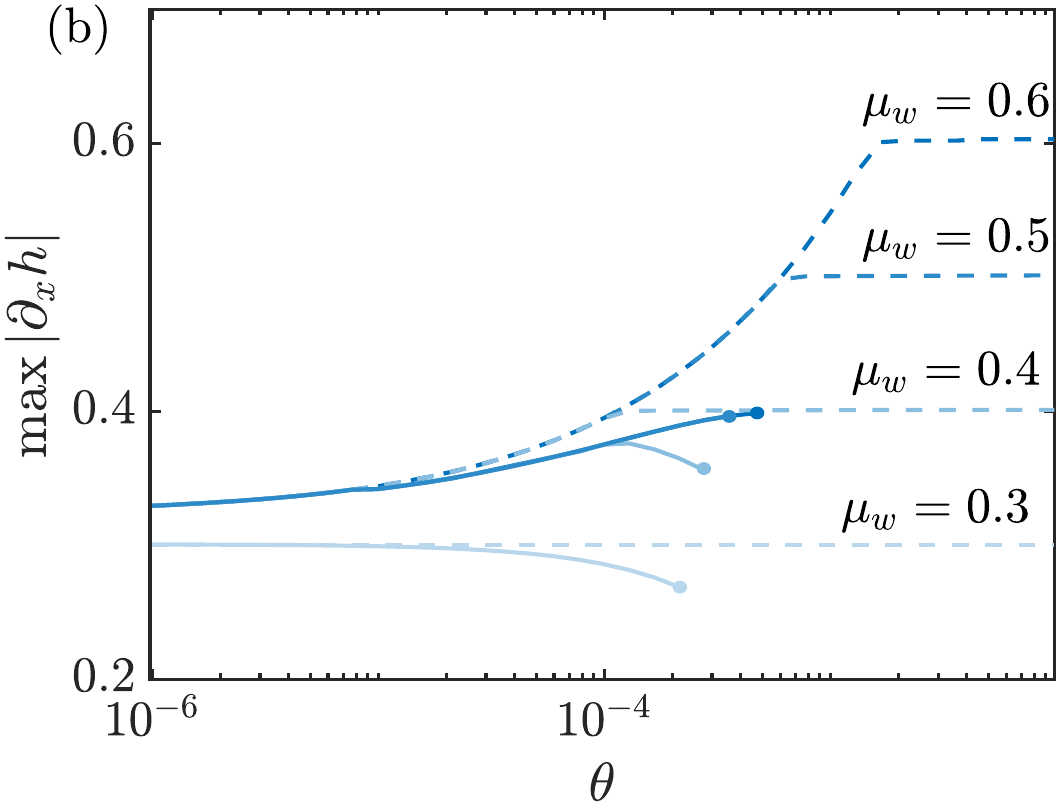}\\
	\includegraphics[width=0.4\linewidth]{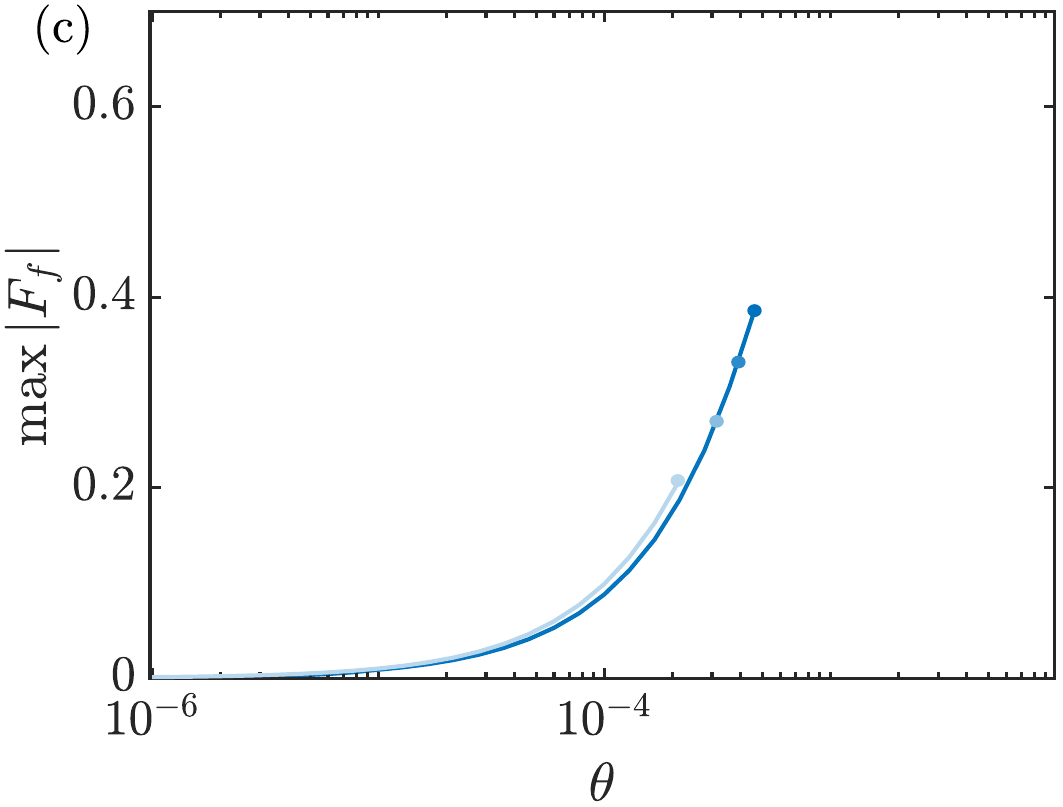}
	\includegraphics[width=0.4\linewidth]{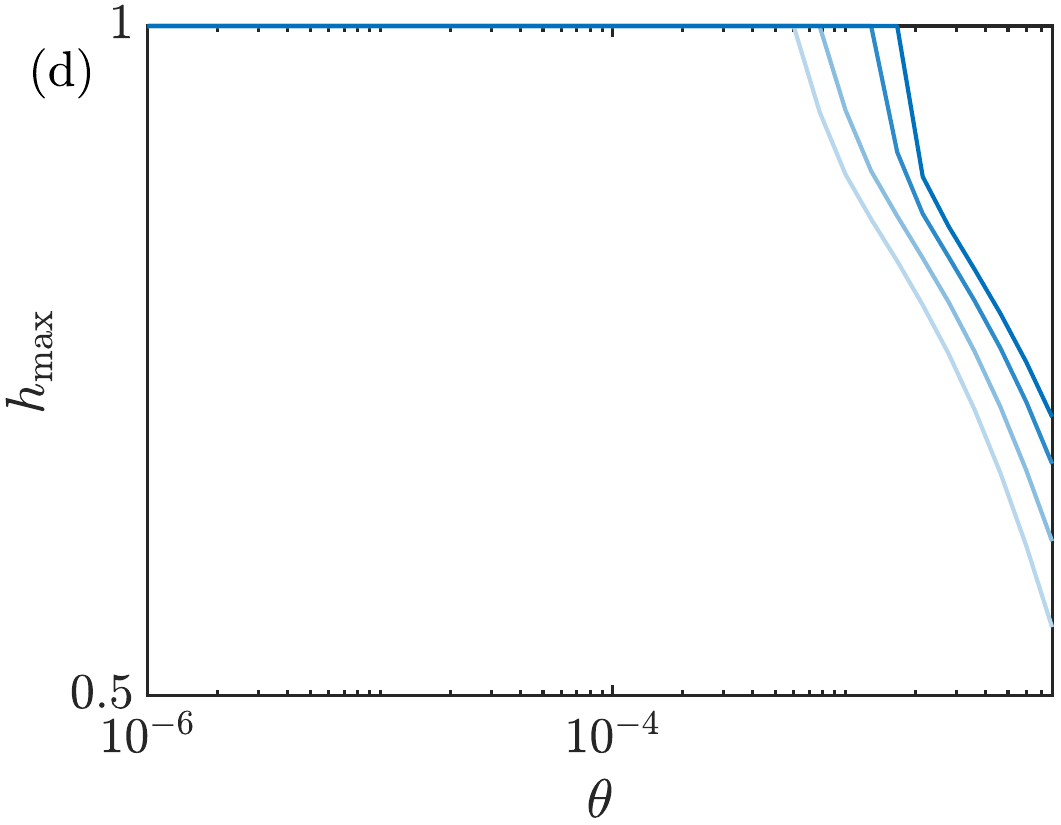}
	\caption{(a) The length of the transition zone, (b) the maximum magnitude of the slope of $h$, (c) the maximum of the fluid force determined by equation \eqref{eq:FluidForce}, and (d) the maximum value of $h$ over the duration of a simulation. The dots at which the various lines terminate denote the largest value of $\theta$ at which the numerical solution converged to a steady travelling wave. The inset in (a)~indicates how the impact of permeability $K$ on the value $\theta = \theta^*$ at which there ceases to be a steady travelling wave for $\mu_w = 0.5$. The dashed lines denote the case where the fluid phase is ignored by solving equations \eqref{eq:EvolutionH}--\eqref{eq:ConservationOfMass} with $\p_x P_f = \sigma_{xz}'(z = h) = 0$. \label{fig:lambdaViscous} }
\end{figure}

This behaviour can be clearly seen in figure~\ref{fig:lambdaViscous}(a), which shows the steady-state length of the transition zone $\lambda$ as a function of $\theta$. The transition zone length gently decreases for small $\theta$ in line with the fluid-free case (dashed lines), but increases sharply for larger $\theta$ as the resistance from the fluid layer starts to have an impact. These lines terminate at a critical value of $\theta$, beyond which computations never evolve to a steady travelling-wave solution. The transition point varies with the wall friction $\mu_w$, but is also strongly influenced by the permeability of the grains (inset of figure \ref{fig:lambdaViscous}a), with higher permeability increasing the value of $\theta$ beyond which no steady solution is attained. Physically, the ability for the upper fluid to flow through the grain pack aides the displacement (via drainage) of that layer, and so grains with lower resistance to drainage (higher permeability) have a higher critical $\theta$. 

The corresponding maximum slope of the free surface (figure~\ref{fig:lambdaViscous}b) similarly peels off from the fluid-free case as $\theta$ is increased, which also reflects the elongation of the transition region. Figure~\ref{fig:lambdaViscous}(c) shows the maximum value of the fluid force in the steady travelling-wave solution, which gives a clear indication of the impact of the fluid layer above the grains: the force is negligible when $\theta$ is sufficiently small, but increases significantly when the fluid-free and full results start to deviate in the first two panels. Note that the force-free (\textit{i.e.}, fluid-free) results in figure~\ref{fig:lambdaViscous} follow much the same trend as the `dry' results in figure~\ref{fig:lambdaDry}, although there are some small quantitative differences from imposing $\mu(I_v)$ instead of $\mu(I)$.  

Figure~\ref{fig:lambdaViscous}(d) shows the maximum height ever reached by any of the grains, which is $1$ (i.e. the height of the channel) if bridging occurs. Evidently the grains never bridge the gap at all if $\theta$ is sufficiently large. However, comparison with the other panels shows that this transition from bridging to no bridging occurs at a notably larger value of $\theta$ than the point at which a steady transition zone is no longer observed: there is evidently a range of $\theta$ for which the grains do bridge the gap but the transition zone does not attain a steady travelling wave. 

\begin{figure}
    \centering
    \includegraphics[width=0.95\linewidth]{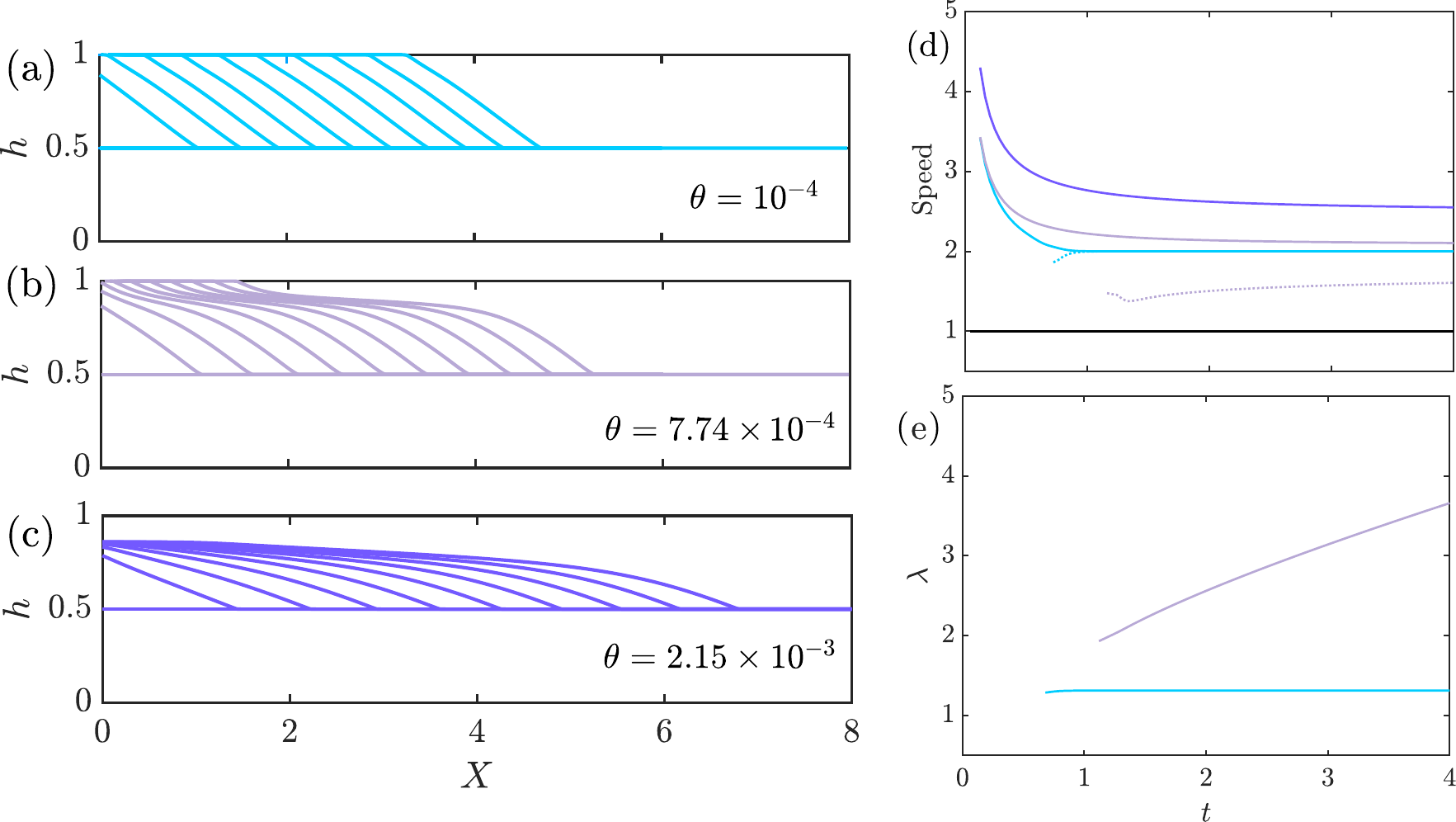}
    \caption{(a)--(c) Evolution of the thickness of the granular layer from the viscous model with $\mu_w =0.5$ for increasing $\theta$ (top to bottom) at equally spaces times in $0\leq{}t<4$. (d)~The associated speed of the leading and trailing edges of the transition region, where these edges exist. The leading and trailing edges are the maximum values of $X$ where $h = h_0$ (solid lines) and where $h = 1$ (dotted lines), respectively. Both speeds exceed the speed of the piston (horizontal black line) in all cases. The two edges converge to the same speed for $\theta=10^{-4}$ (teal) and to different speeds for $\theta=7.74\times{}10^{-4}$ (light purple). The transition region has no trailing edge for the largest value of $\theta$, which never develops a bridged region (dark purple). (e)~Time evolution of the length $\lambda$ of the transition region, defined as the distance between the leading and trailing edges of the transition region. Colours in (d,e) correspond to the different cases in (a--c); the case in (c) never bridges the gap, such that $\lambda$ is not defined. \label{fig:largeTheta} }
\end{figure}

To further explore this behaviour, we show in figure~\ref{fig:largeTheta} details of how $h$ evolves for three different values of $\theta$. The first (figure~\ref{fig:largeTheta}a) attains a steadily propagating transition zone; the second (figure~\ref{fig:largeTheta}b) bridges the gap, but leaves a transition zone that grows steadily longer; and the third (figure~\ref{fig:largeTheta}c) never bridges the gap. Panels (d) and (e) show measurements of the speed of the front and back of the transition zone, and its length, which confirm this behaviour: beyond a certain value of $\theta$ the back of the transition region moves more slowly than the front and the transition region inexorably lengthens over time; beyond a larger value of $\theta$, the grains are unable to bridge the gap and the upper fluid layer reaches all the way back to the piston.

\section{Comparison with Experiments} \label{sec:Experiments}

We present a number of laboratory experiments conducted in a horizontal glass tube with a square cross-section of internal side height $b = 3$ mm and width $W = 3$ mm, and a piston consisting of a steel rod with a moulded gasket (Vytaflex 60). For each experiment, the tube is initially filled with a liquid before the partial loading of wet soda lime glass beads (Honite Blast Media) with $\rho_s = 2480$ kg$\cdot$m$^{-3}$ sieved to size range $150<d <200$~$\mu$m. The granular material was fed via a syringe into the tube, angled at approximately 45 degrees, and allowed to fall due to gravity until forming a layer of height $h_0 = 0.75 \pm 0.15$ mm. The piston was driven into the stationary tube by a syringe pump (Harvard PHD Ultra) at speeds ranging from $1.85 \times 10^{-2}$ to $2.78$ mm/s. Two interstitial liquids were used: water and a solution of 70wt\% glycerol (Fisher Scientific) in water ($\rho_f = 1181$ kg$\cdot$m$^{-3}$ and $\eta_f = 2.25 \times 10^{-2}$ Pa$\cdot$s).  Experiments were captured with a Nikon Z6II camera and 105 mm macro lens. Following previous models of dense granular transport \citep{Aussillous2013,Pailha2009}, we estimate the permeability of the grains via the Kozeny–Carman formula to be $k = 4.5 \times 10^{-5}$ mm$^2$. Further, $\mu_s$ and $\mu_w$ are taken to be 0.48 and 0.53, which we estimated from the angle of repose and the angle at which the beads slide over a flat surface, respectively. The parameters $\mu_2$ and $I_0$ are more difficult to determine without more advanced rheometry, so we use the values reported by \citet{Boyer2011} ($\mu_2 = 0.7$, $I_0 = 0.005$). Over the range of these experiments, the square of the macroscopic Froude number $\mathcal{F}^2$ is never greater than $O(10^{-3})$, suggesting that it is reasonable to neglect the inertia of the grain--fluid mixture. The maximum Reynolds number in the fluid layer $\mathrm{Re}_f$ is $O(1)$, suggesting that fluid inertia should start to play a role in the fastest experiments.

\begin{figure}
	\centering
	\includegraphics[width=0.32\linewidth]{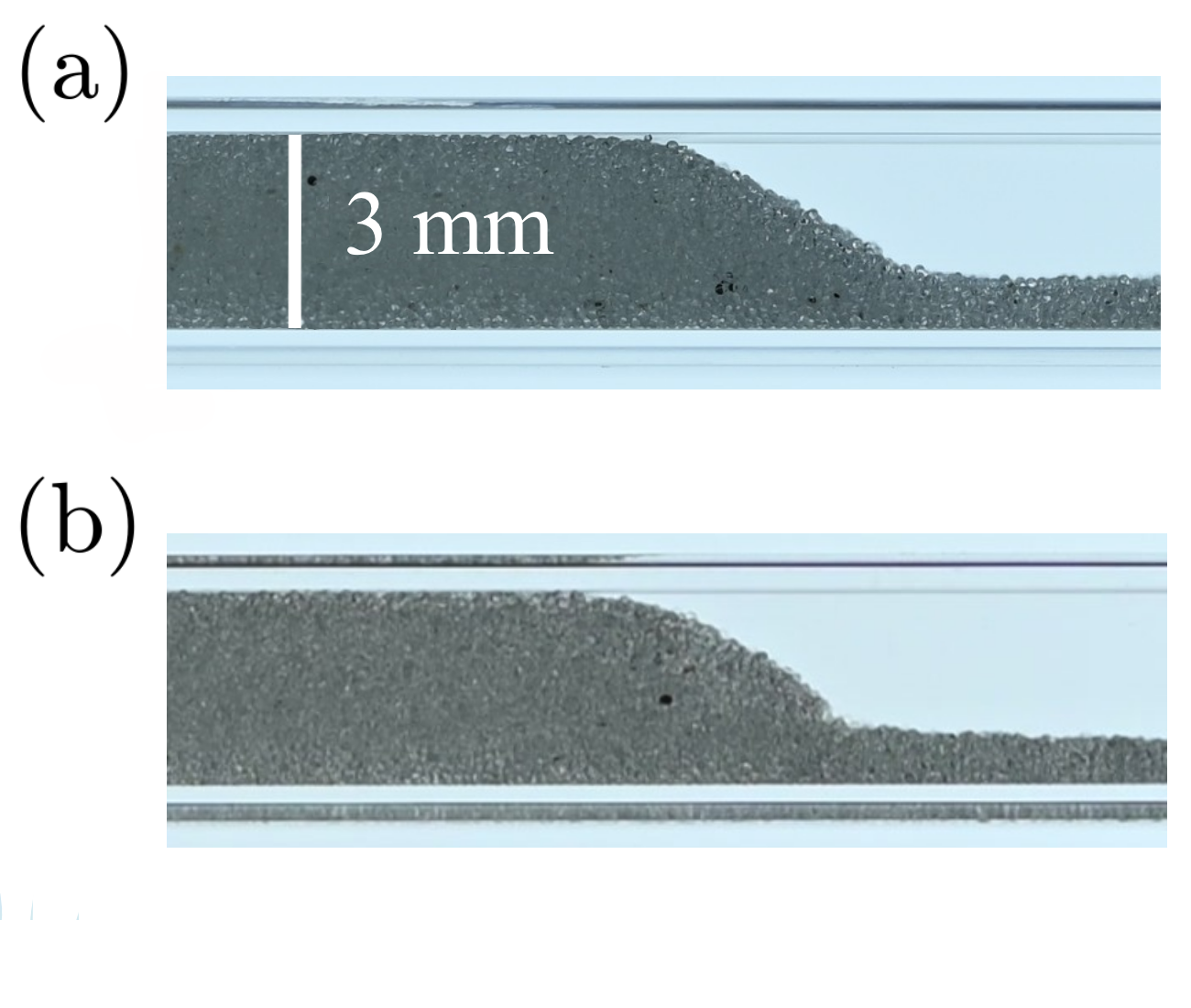}
	\includegraphics[width=0.32\linewidth]{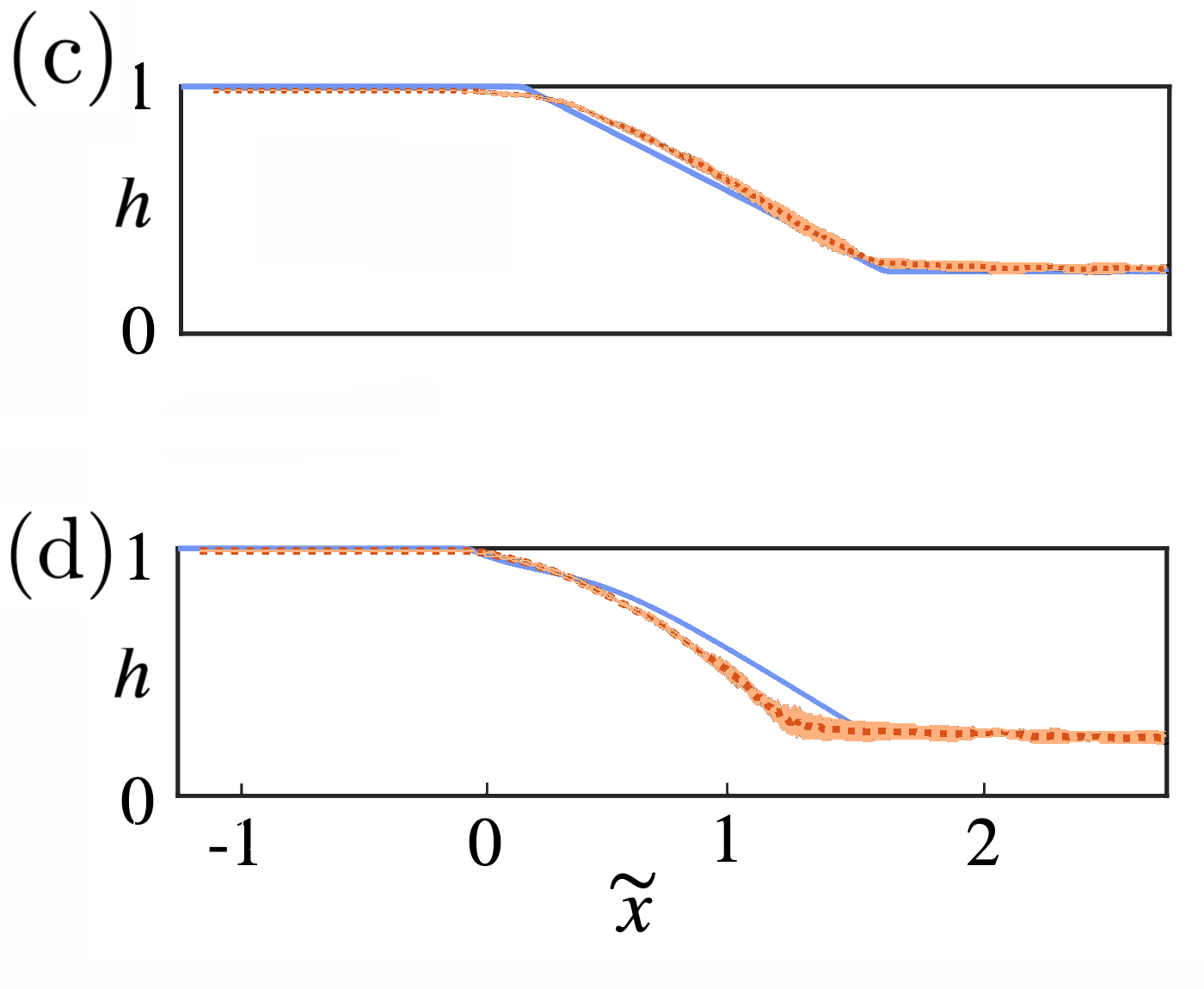}
    \includegraphics[width=0.34\linewidth]{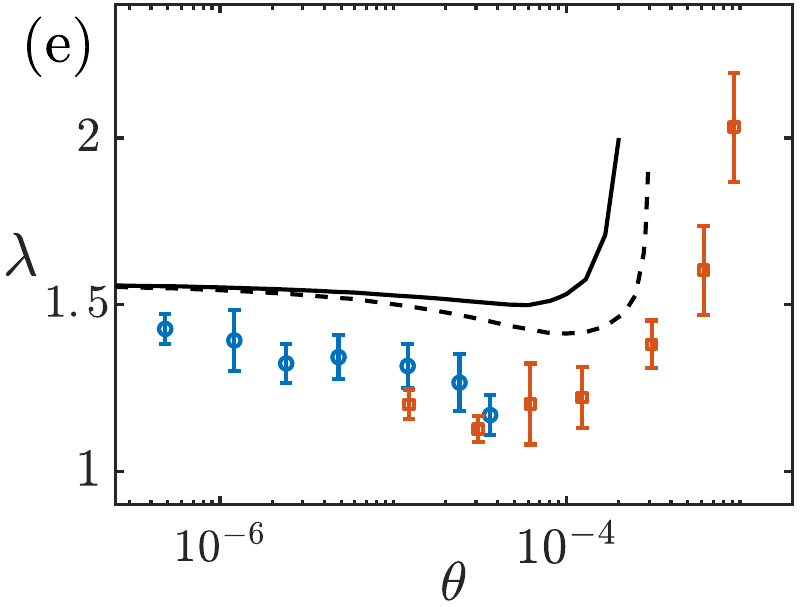}
	\caption{Panels (a) and (b) show experimental snapshots of the travelling-wave state for (a)~$\theta = 4.83 \times 10^{-7}$ (slow pushing) and (b)~$1.23 \times 10^{-4}$ (fast pushing). Panels (c) and (d) show the thickness of the granular layer from the experiments (red) and} from corresponding simulations (without side walls; blue). For the experiments, dotted curves are the mean of three repeats and the light red bands denote one standard deviation above and below the mean.
	(e)~The non-dimensional length of the transition region as a function of $\theta$ from experiments (symbols) and simulations (curves). Blue and red dots denote experiments performed with water and with a solution of 70 wt\% glycerol in water, respectively. Solid and dashed black curves denote simulations performed without and with the effect of side walls, respectively. Simulations are computed with $K = 5 \times 10^{-6}$, $\mu_w = 0.53$, $\mu_s = 0.48$, $I_0 = 0.005$, and $h_0 = 0.25$. \label{fig:lambda} 
\end{figure}

We compare the height of the granular pile determined from experiments and simulations in Fig.~\ref{fig:lambda} for a relatively small [(a) and (c)] and large [(b) and (d)] value of $\theta$, corresponding to relatively slow and fast piston velocity, respectively. There is broad qualitative agreement between the experiments and simulations. That is, experiments indicate that both the length and shape of the transition region is $\theta$-dependent. For low $\theta$, the granular pile appears to have a roughly linear profile, while for a larger value of $\theta$, $h$ appears more non-linear. In Figure~\ref{fig:lambda}(e), we compare the length of the transition zone, which is computed manually from experimental images using the processing software ImageJ, with steady-state measurements of at least three experiments being averaged. Again, both experiments and simulations reproduce the same key behaviours. Specifically, $\lambda$ is non-monotonic in $\theta$, where $\lambda$ decreases due to the balance of frictional forces between grains and walls, before sharply increasing due to the increasing viscous forces imposed on the grains.  We do note however that the experiments measure a lower value of $\lambda$ compared to the model. One possible reason for this discrepancy is that the influence of the sides of the tube is not included in the model.  As such, we extend our model to account for frictional forces due to the sides of the tube, with details being provided in Appendix~\ref{sec:SideWalls}. Note that this extended model includes side-wall friction only under the assumption that there is negligible variation in the flow in the horizontal ($y$) direction, and this assumption must break down if the side-wall friction is sufficiently strong. Figure~\ref{fig:lambda} shows that the inclusion of side-wall friction (dashed line) into the model improves the agreement with experiments, suggesting that such friction has a significant effect. A further possible reason for the discrepency between theory and experiments here could be inaccuracy in the prediction of the permeability of the grains. The results in the inset of figure~\ref{fig:lambdaViscous}(a) showed that a higher permeability leads to a higher value of $\theta$ beyond which no steady $\lambda$ exists, and so we may also be slightly underestimating the permeability of the grain pack here.

\begin{figure}
	\centering
	\includegraphics[width=0.16\linewidth]{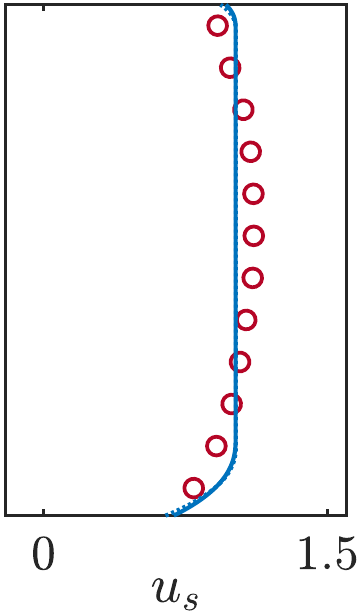}
	\includegraphics[width=0.16\linewidth]{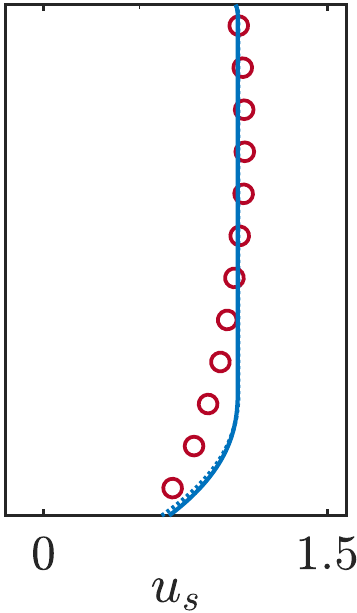}
	\includegraphics[width=0.16\linewidth]{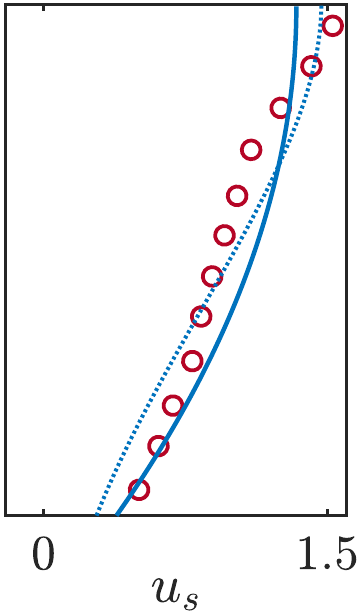}
	\includegraphics[width=0.16\linewidth]{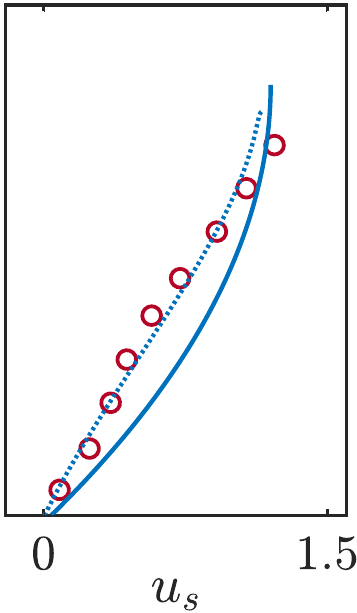}
	\includegraphics[width=0.16\linewidth]{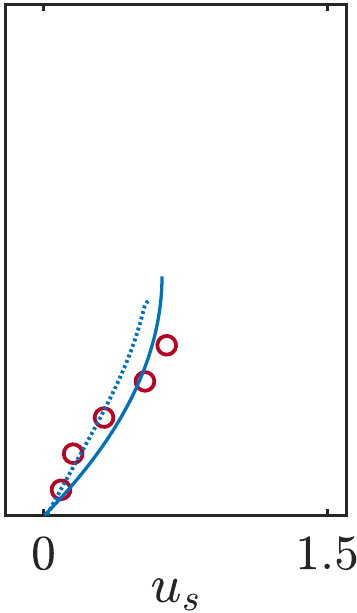}
	\includegraphics[width=0.16\linewidth]{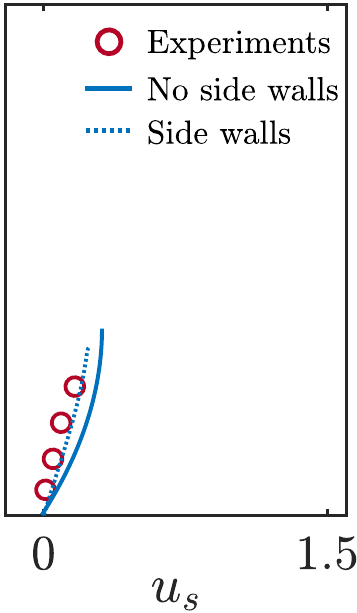}\\
	$\tilde{x} = -1$ \hspace{1.2cm} $\tilde{x} = -0.5$ \hspace{1.2cm} $\tilde{x} = 0$ \hspace{1.2cm} $\tilde{x} = 0.5$ \hspace{1.2cm} $\tilde{x} = 1$ \hspace{1.2cm} $\tilde{x} = 1.5$
	\caption{Slices of the non-dimensional solid velocity profile in the lab frame with $\theta = 1.23 \times 10^{-5}$ from the model (blue) and experiments (red dots) taken at (left to right) $\tilde{x} = -1$, $-0.5$, $\ldots$,  $1.5$. The dotted and solid lines denote profiles determined from the model with and without the effects of the side walls, respectively. }
	\label{fig:VelocitySlices}
\end{figure}

We next compare the solid velocity profiles from experiments with simulations for a `moderate' value of $\theta$, shown in  Fig.~\ref{fig:VelocitySlices}. We measure the solid velocity in the experiments from from side-wall observations, noting that observations from above suggest that the flow field is fairly uniform across the cross-section. We position-correct videos to track the start of the transition region and then compute the steady velocity field using particle image velocimetry (PIV), time-averaging over several frames~\citep{Thielicke2014}. We find that simulations and experiments are in reasonable agreement with each other. In the transition region, $\tilde{x} \ge 0$, experiments show that the velocity profile is monotonically increasing in the vertical direction, and the shear stress at $z = 0$ is sufficient for the grains to slide over the base, which is consistent with the model. Further, for a smaller $\theta$ (not shown), grains in contact with the bottom of the tube are stationary, again consistent with simulations. We do note, however, that experiments appear to produce profiles where $\p_z^2 u_s > 0$, while the model produces profiles where $\p_z^2 u_s < 0$ for all values of $\theta$.  Incorporating side walls into the model (dashed lines) can produce profiles where  $\p_z^2 u_s < 0$, suggesting again that the presence of the sides of the tube has a significant influence on the dynamics.

Regarding the velocity profile in the bridged region, recall that our model exhibits a piecewise velocity profile of the form of eq.~\eqref{eq:PiecewiseVelocity} illustrated in Figure~\ref{fig:sketchshearbridge}.  Figure~\ref{fig:VelocitySlices} shows that qualitatively, experiments and simulations produce similar behaviour, where grains move slower near the top and bottom walls and faster in the interior, away from the walls. One notable difference between the two is that, unlike the simulations, the experiments do not exhibit a yielded profile in the interior. Unlike in the transition region, accounting for side walls does not appear to have a qualitatively or quantitatively important impact on the velocity profile, and the two scenarios (side walls and no side walls) appear to converge moving to the left, away from the transition region. However, side walls do have an important impact on the bulldozing force. As discussed in \S\ref{sec:DryBridgedZone}, we find that without side walls $|\p_x P_s| \sim 2\mu_w P_s$ in the bridged region; with side walls, the growth rate of the normal stress in the bridged region effectively doubles, such that $|\p_x P_s| \sim 4\mu_w P_s$.  We further note that the vertical velocity of the grains (not shown) from experiments is several orders of magnitude smaller than the horizontal component across the entire domain, supporting our assumption of a lubrication-type flow ($w_s \ll u_s$; \S \ref{sec:LowerLayer}).

\section{Conclusion} \label{sec:Discussion}

We have presented a continuum model for the bulldozing of a fluid-saturated granular material in a confined geometry. We derived this model by framing the problem as a depth-averaged system of coupled thin films: an incompressible Newtonian fluid overlying a deformable, porous granular layer. To model intra-granular friction, we imposed either the inertial $\mu(I)$ rheology (`dry' case) or the viscous $\mu(I_v)$ rheology (`wet' case), depending on the relative importance of fluid viscosity --- as quantified by the value of the Stokes number. We modelled contact with the bottom and top walls as simple Coulomb friction, such that the shear stress at the wall must be sufficiently large for the grains to slide. We neglect macroscopic inertial forces in both the fluid layer and the granular suspension. By solving our model numerically, we investigated the motion of the grains in the region where the granular layer has bridged the gap between the plates and in the transition zone between the bridged region and the undisturbed far-field. We also provide a comparison of our model with experiments in \S\ref{sec:Experiments}.

In \S\ref{sec:DryRegion}, we considered the `dry' case in which the fluid is negligible. We found that, in the bridged region ($h = b$), the grains exhibit a non-monotonic velocity profile comprising a plugged region ($\p_z u_s = 0$) sandwiched between two yielded regions ($|\p_z u_s| > 0$). This profile comes from a balance between the rate-dependent frictional intra-granular stresses with sliding friction at the walls. Further, we find that the normal stress at the piston grows exponentially with the length of the bridged region, as expected from previous work. In \S\ref{sec:DryTransitionZone}, we considered the behaviour of the grains in the wedge-shaped transition zone. We found that the shape of the transition zone depends on the Froude number, generally decreasing in length with increasing $\Fr$ due to the increasing resistance to internal rearrangement as the shear rate increases. As in the bridged region, the specific velocity profile and height of the granular layer are determined by a balance between intra-granular friction and wall friction.

In \S\ref{sec:ViscousModel}, we considered the `wet' case in which fluid viscosity plays an increasingly important role at faster rates. We found that, for a sufficiently small Shields number $\theta$, the impact of the fluid is negligible and solutions exhibit the same qualitative behaviour as the `dry' model. Specifically, the transition zone tends toward a triangular wedge with a slope $\mu_s$ (\textit{i.e.}, the static angle of repose) for $\theta \ll 1$. This wedge shortens as $\theta$ increases, with the slope steepening toward $\mu_w$ (provided that $\mu_w > \mu_s$). As $\theta$ increases further, the transition zone instead eventually begins to lengthen and becomes less triangular due to the increasingly important role of fluid viscosity, unlike in the dry case. For sufficiently large $\theta$, our model predicts that the transition region has no steady shape; rather, it lengthens continually as viscous shear from the overlying fluid acts to flatten the granular free surface. For even larger $\theta$, viscous shear is sufficiently strong that the granular layer is unable to bridge the gap.

This non-monotonic variation of the length of the transition region with $\theta$ in the `wet' case is consistent with experiments presented in \S\ref{sec:Experiments}, which also show a gentle decrease and then a steep increase in $\lambda$ as $\theta$ increases. For low $\theta$, both model and experiments show that the grains in the transition zone will exhibit no-slip on the bottom wall, with a horizontal velocity that increases monotonically in $z$. For larger $\theta$, the shear stress at the base becomes sufficiently large such that the grains can slide along the bottom wall. 

Although the model presented here produces reasonable qualitative agreement with experiments, several improvements are possible. For example, the model provides a somewhat better quantitative match with experiments when friction with the side walls is included. In \S\ref{sec:ModelDevelopment}, we assumed that the solid fraction of granular layer is constant; however, in reality, granular materials are known to compact or dilate under shear. This behaviour can be incorporated by introducing additional constitutive laws for $\phi(I)$ for the inertial regime \citep{Forterre2008} or $\phi(I_v)$ for the viscous regime \citep{Boyer2011}.  Similarly, our model ignores inertial effects at the continuum scale, which must become more important at higher shear rates and could, for example, enable the granular layer to overcome viscous resistance from the the fluid and bridge the gap for large $\theta$ .Our experiments at the highest values of $\theta$ had $\mathcal{F}^2 \sim 10^{-3}$ but a fluid Reynolds number $\mathrm{Re}_f\sim 1$, suggesting that inertia in the fluid layer is likely to have more of an effect on the dynamics here than inertia in the granular mixture.
Finally, although our model is based on a lubrication approximation, the flow may not be predominantly horizontal in all cases; as a result, secondary recirculating flow or non-trivial normal stresses could become important and are not captured in this modelling framework.

As discussed in \S\ref{sec:Introduction}, one motivation for this work was to provide insight into the behaviour of a granular material that is being bulldozed by a fluid-fluid interface, such as during the injection of a gas into a hydrophilic, water-saturated packing. As long as capillary forces are sufficient to prevent the gas from invading the pore space of the packing, this scenario can be thought of as the bulldozing of grains by a deformable piston. Experiments in both capillary tubes \citep{Dumazer2016,Thorens2023} and Hele-Shaw cells \citep{Sandnes2011} show the formation of a variety of patterns resulting from the balance between frictional and capillary forces. With this in mind, an avenue for future work is to extend the two-phase model presented here to incorporate an additional fluid that is non-wetting to the grains.

\backsection[Funding]{This work was supported by the European Research Council (ERC) under the European Union's Horizon 2020 Programme [Grant No. 805469] and by the UK Engineering and Physical Sciences Research Council (EPSRC) [Grant No. EP/S034587/1 and EP/X028771/1].}

\backsection[Declaration of interests]{The authors report no conflict of interest.}

\bibliographystyle{jfm}
\bibliography{MyBibFile}

@book{Andreotti2013,
	title={Granular media: between fluid and solid},
	author={Andreotti, B. and Forterre, Y. and Pouliquen, O.},
	year={2013},
	publisher={Cambridge University Press}
}

@article{putz,
title = {On the lubrication paradox and the use of regularisation methods for lubrication flows},
journal = {Journal of Non-Newtonian Fluid Mechanics},
volume = {163},
number = {1},
pages = {62-77},
year = {2009},
issn = {0377-0257},
doi = {https://doi.org/10.1016/j.jnnfm.2009.06.006},
url = {https://www.sciencedirect.com/science/article/pii/S0377025709001396},
author = {A. Putz and I.A. Frigaard and D.M. Martinez},
}

@article{Barker_wellposed1, 
    title={Well-posed and ill-posed behaviour of the ${\it\mu}(I)$ -rheology for granular flow}, 
    volume={779}, 
    DOI={10.1017/jfm.2015.412}, 
    journal={Journal of Fluid Mechanics}, 
    author={Barker, T. and Schaeffer, D. G. and Bohorquez, P. and Gray, J. M. N. T.}, 
    year={2015}, 
    pages={794–818}
}

@article{Barker_wellposed_susp, 
    title={Well-posedness and ill-posedness of single-phase models for suspensions}, 
    volume={954}, 
    DOI={10.1017/jfm.2022.1004}, 
    journal={Journal of Fluid Mechanics}, 
    author={Barker, T. and Gray, J. M. N. T. and Schaeffer, D. G. and Shearer, M.}, 
    year={2023}, 
    pages={A17}
}

@article{barker_wellposed_comp,
    author = {Barker, T. and Schaeffer, D. G. and Shearer, M. and Gray, J. M. N. T.},
    title = {Well-posed continuum equations for granular flow with compressibility and ${\it\mu}(I)$-rheology},
    journal = {Proceedings of the Royal Society A: Mathematical, Physical and Engineering Sciences},
    volume = {473},
    pages = {20160846},
    year = {2017},
    issn = {1364-5021},
    doi = {10.1098/rspa.2016.0846},
    url = {https://doi.org/10.1098/rspa.2016.0846},
    eprint = {https://royalsocietypublishing.org/rspa/article-pdf/doi/10.1098/rspa.2016.0846/364225/rspa.2016.0846.pdf},
}

@article{Artoni2015,
	title={Effective wall friction in wall-bounded {3D} dense granular flows},
	author={Artoni, R. and Richard, P.},
	journal={Phys.~Rev.~Lett.},
	volume={115},
	pages={158001},
	year={2015},
	DOI={0.1103/PhysRevLett.115.158001}
}

@article{viscop,
   author = "Balmforth, Neil J. and Frigaard, Ian A. and Ovarlez, Guillaume",
   title = "Yielding to Stress: Recent Developments in Viscoplastic Fluid Mechanics", 
   journal= "Annual Review of Fluid Mechanics",
   year = "2014",
   volume = "46",
   number = "Volume 46, 2014",
   pages = "121-146",
   doi = "https://doi.org/10.1146/annurev-fluid-010313-141424",
   url = "https://www.annualreviews.org/content/journals/10.1146/annurev-fluid-010313-141424",
   publisher = "Annual Reviews",
   issn = "1545-4479",
   type = "Journal Article",
  }

@article{Aussillous2013,
	title={Investigation of the mobile granular layer in bedload transport by laminar shearing flows},
	author={Aussillous, P. and Chauchat, J. and Pailha, M. and M{\'e}dale, M. and Guazzelli, {\'E}.},
	journal={J. Fluid Mech.},
	volume={736},
	pages={594--615},
	year={2013},
	DOI={10.1017/jfm.2013.546}
}

@article{Baumgarten2019,
	title={A general fluid--sediment mixture model and constitutive theory validated in many flow regimes},
	author={Baumgarten, A. S. and Kamrin, K.},
	journal={J. Fluid Mech.},
	volume={861},
	pages={721--764},
	year={2019},
	DOI={10.1017/jfm.2018.914}
}

@article{Beavers1967,
  title={Boundary conditions at a naturally permeable wall},
  author={Beavers, G. S. and Joseph, D. D.},
  journal={J. Fluid Mech.},
  volume={30},
  pages={197--207},
  year={1967},
  publisher={Cambridge University Press}
}

@article{Bouchut2016,
	title={A two-phase two-layer model for fluidized granular flows with dilatancy effects},
	author={Bouchut, F. and Fern{\'a}ndez-Nieto, E. D. and Mangeney, . and Narbona-Reina, G.},
	journal={J. Fluid Mech.},
	volume={801},
	pages={166--221},
	year={2016}
}

@article{Boyer2011,
	title={Unifying suspension and granular rheology},
	author={Boyer, F. and Guazzelli, {\'E}. and Pouliquen, O.},
	journal={Phys.~Rev.~Lett.},
	volume={107},
	pages={188301},
	year={2011},
	DOI={10.1103/PhysRevLett.107.188301}
}

@article{Cassar2005,
	title={Submarine granular flows down inclined planes},
	author={Cassar, C. and Nicolas, M. and Pouliquen, O.},
	journal={Phys.~Fluids},
	volume={17},
	pages={103301},
	year={2005},
	DOI={10.1063/1.2069864}
}

@article{DaCruz2005,
	title={Rheophysics of dense granular materials: Discrete simulation of plane shear flows},
	author={Da Cruz, F. and Emam, S. and Prochnow M. and Roux, J. N. and Chevoir, F.},
	journal={Phys.~Rev.~E},
	volume={16},
	pages={021309},
	year={2005},
	DOI={10.1103/PhysRevE.72.021309}
}

@article{Doppler2007,
	title={Relaxation dynamics of water-immersed granular avalanches},
	author={Doppler, D. and Gondret, P. and Loiseleux, T. and Meyer, S. and Rabaud, M.},
	journal={J. Fluid Mech.},
	volume={577},
	pages={161--181},
	year={2007},
	DOI={10.1017/S0022112007004697}
}

@article{Dorostkar2017,
	title={On the micromechanics of slip events in sheared, fluid-saturated fault gouge},
	author={Dorostkar, O. and Guyer, R. A. and Johnson, P. A. and Marone, C. and Carmeliet, J.},
	journal={Geophys. Res. Lett.},
	volume={44},
	pages={6101--6108},
	year={2017}
}

@article{Dumazer2016,
	title={Frictional fluid dynamics and plug formation in multiphase millifluidic flow},
	author={Dumazer, G. and Sandnes, B. and Ayaz, M. and M{\aa}l{\o}y, K. J. and Flekk{\o}y, E. G.},
	journal={Phys.~Rev.~Lett.},
	volume={117},
	pages={028002},
	year={2016},
	DOI={10.1103/PhysRevLett.117.028002}
}

@article{Dumazer2020,
	title={Capillary bulldozing of sedimented granular material confined in a millifluidic tube},
	author={Dumazer, G. and Sandnes, B. and M{\aa}l{\o}y, K. J. and Flekk{\o}y, E. G.},
	journal={Phys.~Rev.~Fluids},
	volume={5},
	pages={034309 },
	year={2020},
	DOI={10.1103/PhysRevFluids.5.034309}
}

@article{Eriksen2015,
	title={Numerical approach to frictional fingers},
	author={Eriksen, J. A. and Toussaint, R. and M{\aa}l{\o}, K. J. and Flekk{\o}y, E. and Sandnes, B.},
	journal={Phys.~Rev.~E},
	volume={92},
	pages={032203},
	year={2015},
	DOI={10.1103/PhysRevE.92.032203}
}

@article{Eriksen2015b,
	title={Bubbles breaking the wall: Two-dimensional stress and stability analysis},
	author={Eriksen, J. A. and Sandnes, B. and Marks, B. and Toussaint, R.},
	journal={Phys.~Rev.~E},
	volume={91},
	pages={052204},
	year={2015},
	DOI={10.1103/PhysRevE.91.052204}
}

@article{Eriksen2018,
	title={Pattern formation of frictional fingers in a gravitational potential},
	author={Eriksen, J. A. and Toussaint, R. and M{\aa}l{\o}y, K. J. and Flekk{\o}y, E. and Galland, O. and Sandnes, B.},
	journal={Phys. Rev. Fluids},
	volume={3},
	pages={013801},
	year={2018},
	DOI={10.1103/PhysRevFluids.3.013801}
}

@article{Forterre2008,
	title={Flows of dense granular media},
	author={Forterre, Y. and Pouliquen, O.},
	journal={Annu. Rev. Fluid Mech.},
	volume={40},
	pages={1--24},
	year={2008},
	DOI={10.1146/annurev.fluid.40.111406.102142}
}

@article{Freund2011,
	title={Cellular flow in a small blood vessel},
	author={Freund, J. B. and Orescanin, M. M.},
	journal={J. Fluid Mech.},
	volume={671},
	pages={466--490},
	year={2011},
	DOI={10.1017/S0022112010005835}
}

@article{Frey2011,
	title={Bedload: a granular phenomenon},
	author={Frey, P. and Church, M.},
	journal={ESPL},
	volume={36},
	pages={58--69},
	year={2011}
}

@article{Hampton1996,
	title={Submarine landslides},
	author={Hampton, M. A. and Homa, J. and Locat, J.},
	journal={Rev. Geophys.},
	volume={34},
	pages={33--59},
	year={1996},
	DOI={10.1029/95RG03287}
}

@article{Houssais2016,
	title={Rheology of sediment transported by a laminar flow},
	author={Houssais, M. and Ortiz, C. P. and Durian, D. J. and Jerolmack, D. J.},
	journal={Phys.~Rev.~E},
	volume={94},
	pages={062609},
	year={2016},
	DOI={10.1103/PhysRevE.94.062609}
}

@article{Iordanoff2004,
	title={Granular lubrication: toward an understanding between kinetic and fluid regime},
	author={Iordanoff, I. and Khonsari, M. M.},
	journal={ASME J.~Tribol.},
	volume={126},
	pages={137--145},
	year={2004},
	DOI={10.1115/1.1633575}
}

@article{Iverson1997,
	title={The physics of debris flows},
	author={Iverson, R. M.},
	journal={Rev.~Geophys.},
	volume={35},
	pages={245--296},
	year={1997},
	DOI={10.1029/97RG00426}
}

@article{Janssen1895,
	title={Versuche uber getreidedruck in silozellen},
	author={Janssen, H. A.},
	journal={Z. ver. deut. Ing.},
	volume={39},
	pages={1045},
	year={1895}
}

@article{Jop2005,
	title={Crucial role of sidewalls in granular surface flows: consequences for the rheology},
	author={Jop, P. and Forterre, Y. and Pouliquen, O.},
	journal={J. Fluid Mech.},
	volume={541},
	pages={167--192},
	year={2005},
	DOI={10.1017/S0022112005005987}
}

@article{Jop2006,
	title={A constitutive law for dense granular flows},
	author={Jop, P. and Forterre, Y. and Pouliquen, O.},
	journal={Nature},
	volume={441},
	pages={727--730},
	year={2006},
	doi={10.1038/nature04801}
}

@article{Knudsen2008,
	title={Granular labyrinth structures in confined geometries},
	author={Knudsen, H. A. and Sandnes, B. and Flekk{\o}y, E. G. and M{\aa}l{\o}y, K. J.},
	journal={Phys.~Rev.~E},
	volume={77},
	pages={021301},
	year={2008},
	DOI={10.1103/PhysRevE.77.021301}
}

@article{Landry2003,
	title={Confined granular packings: structure, stress, and forces},
	author={Landry, L. W. and Grest, G. S. and Silbert, L. E. and Plimpton, S. J.},
	journal={Phys.~Rev.~E},
	volume={67},
	pages={041303},
	year={2003},
	doi={10.1103/PhysRevE.67.041303}
}

@article{Lawal1993,
  title={Extrusion and lubrication flows of viscoplastic fluids with wall slip},
  author={Lawal, A and Kalyon, D M and Yilmazer, U},
  journal={Chem. Eng. Commun.},
  volume={122},
  pages={127--150},
  year={1993},
  publisher={Taylor \& Francis}
}

@article{Legros2002,
	title={The mobility of long-runout landslides},
	author={Legros, F.},
	journal={Eng.~Geol.},
	volume={63},
	pages={301--331},
	year={2002},
	DOI={10.1016/S0013-7952(01)00090-4}
}

@article{LeBarsWorster2006,
	title={Interfacial conditions between a pure fluid and a porous medium: implications for binary alloy solidification},
	author={Le Bars, M. and Worster, M.G.},
	journal={J. Fluid Mech.},
	volume={550},
	pages={149--173},
	year={2006},
	DOI={10.1017/S0022112005007998}
}

@article{Marks2015,
	title={Compaction of granular material inside confined geometries},
	author={Marks, B. and Sandnes, B. and Dumazer, G. and Eriksen, J. A. and M{\aa}l{\o}y, K. J.},
	journal={Front. Phys.},
	volume={3},
	pages={41},
	year={2015},
	DOI={10.3389/fphy.2015.00041}
}

@article{Maurin2016,
	title={Dense granular flow rheology in turbulent bedload transport},
	author={Maurin, R. and Chauchat, J. and Frey, P.},
	journal={J. Fluid Mech.},
	volume={804},
	pages={490--512},
	year={2016},
	DOI={10.1017/jfm.2016.520}
}

@article{Midi2004,
	title={On dense granular flows},
	author={Midi, G. D. R.},
	journal={Eur.~Phys.~J.~E},
	volume={14},
	pages={341--365},
	year={2004},
	DOI={10.1140/epje/i2003-10153-0}
}

@article{Morgan2020,
	title={Self-similar velocity profiles and mass transport of grains carried by fluid through a confined channel},
	author={M. L. Morgan and D. W. James and A. R. Barron and B. Sandnes},
	journal={Phys. Fluids},
	volume={32},
	pages={113309},
	year={2020},
	DOI={10.1063/5.0031155}
}

@article{Ouriemi2009,
	title={Sediment dynamics. Part 1. Bed-load transport by laminar shearing flows},
	author={Ouriemi, M. and Aussillous, P. and Guazzelli, {\'E}.},
	journal={J. Fluid Mech.},
	volume={636},
	pages={295--319},
	year={2009},
	DOI={10.1017/S0022112009007915}
}

@article{Pailha2009,
	title={A two-phase flow description of the initiation of underwater granular avalanches},
	author={Pailha, M. and Pouliquen, O.},
	journal={J. Fluid Mech.},
	volume={633},
	pages={115--135},
	year={2009},
	DOI={10.1017/S0022112009007460}
}

@article{Paterson1981,
	title={{Radial fingering in a Hele Shaw cell}},
	author={Paterson, L.},
	journal={J. Fluid Mech.},
	volume={113},
	pages={513--529},
	year={1981},
	DOI={10.1017/S0022112081003613}
}

@article{Pennestri2016,
	title={Review and comparison of dry friction force models},
	author={Pennestr{\`\i}, E. and Rossi, V. and Salvini, P. and Valentini, P. P.},
	journal={Nonlinear Dyn.},
	volume={83},
	pages={1785--1801},
	year={2016},
	DOI={10.1007/s11071-015-2485-3}
}

@article{Pouliquen2006,
	title={{Flow of dense granular material: towards simple constitutive laws}},
	author={Pouliquen, O. and Cassar, C. and Jop, P. and Forterre, Y. and Nicolas, M.},
	journal={JSTAT},
	volume={2006},
	pages={P07020},
	year={2006},
	DOI={10.1088/1742-5468/2006/07/P07020}
}

@article{Rettinger2022,
	title={{Rheology of mobile sediment beds in laminar shear flow: effects of creep and polydispersity}},
	author={Rettinger, C. and Eibl, S. and R{\"u}de, U. and Vowinckel, B.},
	journal={J. Fluid Mech.},
	volume={932},
	pages={A1},
	year={2022},
	DOI={10.1017/jfm.2021.870}
}

@article{Saffman1958,
	title={{The penetration of a fluid into a porous medium or Hele-Shaw cell containing a more viscous liquid}},
	author={Saffman, P. G. and Taylor, G. I.},
	journal={{Proc. Math. Phys. Eng. Sci.}},
	volume={245},
	pages={312--329},
	year={1958},
	DOI={10.1098/rspa.1958.0085}
}

@article{Sandnes2007,
	title={Labyrinth patterns in confined granular-fluid systems},
	author={Sandnes, B. and Knudsen, H. A. and M{\aa}l{\o}y, K. J. and Flekk{\o}y, E. G.},
	journal={Phys.~Rev.~Lett.},
	volume={99},
	pages={038001},
	year={2007},
	DOI={10.1103/PhysRevLett.99.038001}
}

@article{Sandnes2011,
	title={Patterns and flow in frictional fluid dynamics},
	author={Sandnes, B. and Flekk{\o}y, E. G. and Knudsen, H. A. and M{\aa}l{\o}y, K. J. and See, H.},
	journal={Nat. Commun.},
	volume={2},
	pages={288},
	year={2011},
	DOI={10.1038/ncomms1289}
}

@article{Sauret2014,
	title={Bulldozing of granular material},
	author={Sauret, A. and Balmforth, N. J. and Caulfield, C. P. and McElwaine, J. N.},
	journal={J. Fluid Mech.},
	volume={748},
	pages={143--174},
	year={2014},
	DOI={10.1017/jfm.2014.181}
}

@article{Taberlet2008,
	title={The effect of sidewall friction on dense granular flows},
	author={Taberlet, N. and Richard, P. and Delannay, R.},
	journal={Comput. Math. Appl.},
	volume={55},
	pages={230--234},
	year={2008},
	DOI={10.1016/j.camwa.2007.04.013}
}

@article{Tapia2022,
  title={Viscous to inertial transition in dense granular suspension},
  author={Tapia, F. and Ichihara, M. and Pouliquen, O. and Guazzelli, E},
  journal={Physical Review Letters},
  volume={129},
  pages={078001},
  year={2022},
  DOI={10.1103/PhysRevLett.129.078001}
}

@inproceedings{Terzaghi1936,
	title={The shearing resistance of saturated soils and the angle between the planes of shear},
	author={Terzaghi, K.},
	booktitle={First international conference on soil Mechanics, 1936},
	volume={1},
	pages={54--59},
	year={1936}
}

@article{Thielicke2014,
	author = {Thielicke, W. and Stamhuis, E. J.},
	journal = {Journal of Open Research Software},
	title = {{PIVlab – Towards User-friendly, Affordable and Accurate Digital Particle Image Velocimetry in MATLAB}},
	volume = {2},
	year = {2014}
}

@article{Thorens2023,
  title={Capillary washboarding during slow drainage of a frictional fluid},
  author={Thorens, L. and M{\aa}l{\o}y, K. J. and Flekk{\o}y, E. G. and Sandnes, B. and Bourgoin, M. and Santucci, S.},
  journal={Soft Matter},
  volume={19},
  pages={9369--9378},
  year={2023}
}

@article{Zhang2023,
  title={Frictional fluid instabilities shaped by viscous forces},
  author={Zhang, D. and Campbell, J. M. and Eriksen, J. A. and Flekk{\o}y, E. G. and M{\aa}l{\o}y, K. J. and MacMinn, C. W. and Sandnes, B.},
  journal={Nat. Commun.},
  volume={14},
  pages={3044},
  year={2023},
  publisher={Nature Publishing Group UK London}
}

\appendix

\section{Validation of regularisation} \label{sec:regularisation}

\begin{figure}
    \centering
    \includegraphics[width=0.3\linewidth]{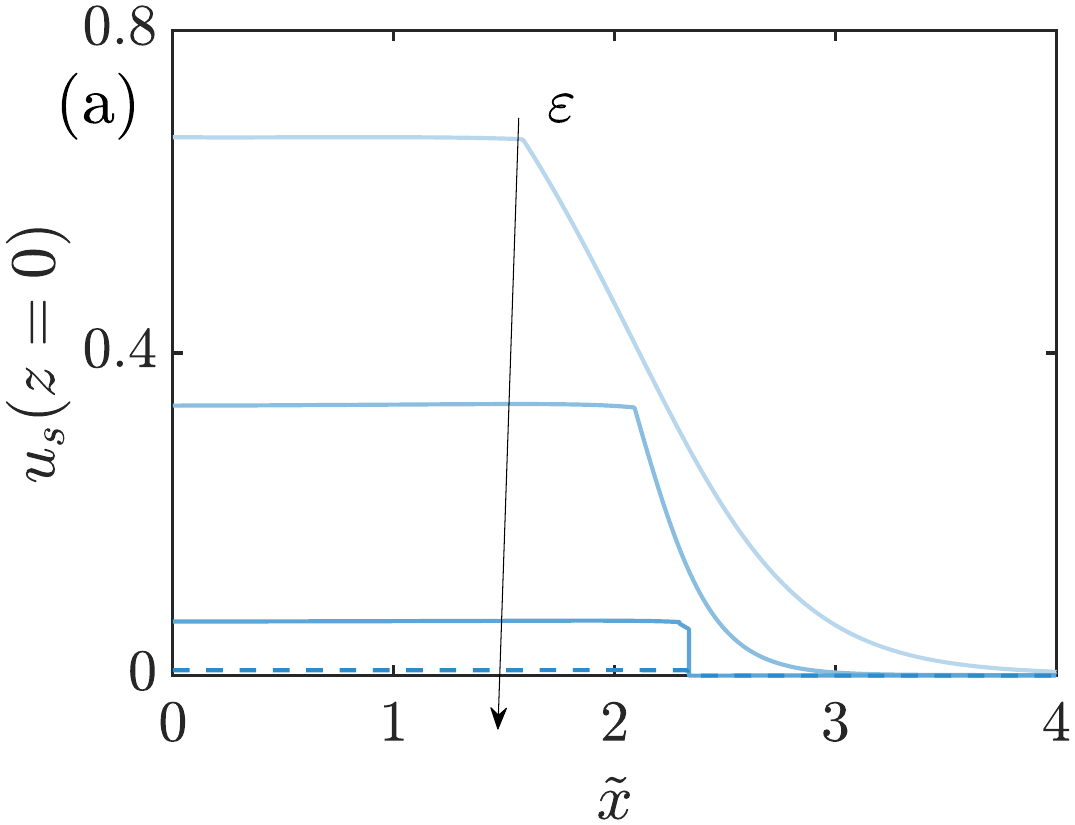}
    \includegraphics[width=0.3\linewidth]{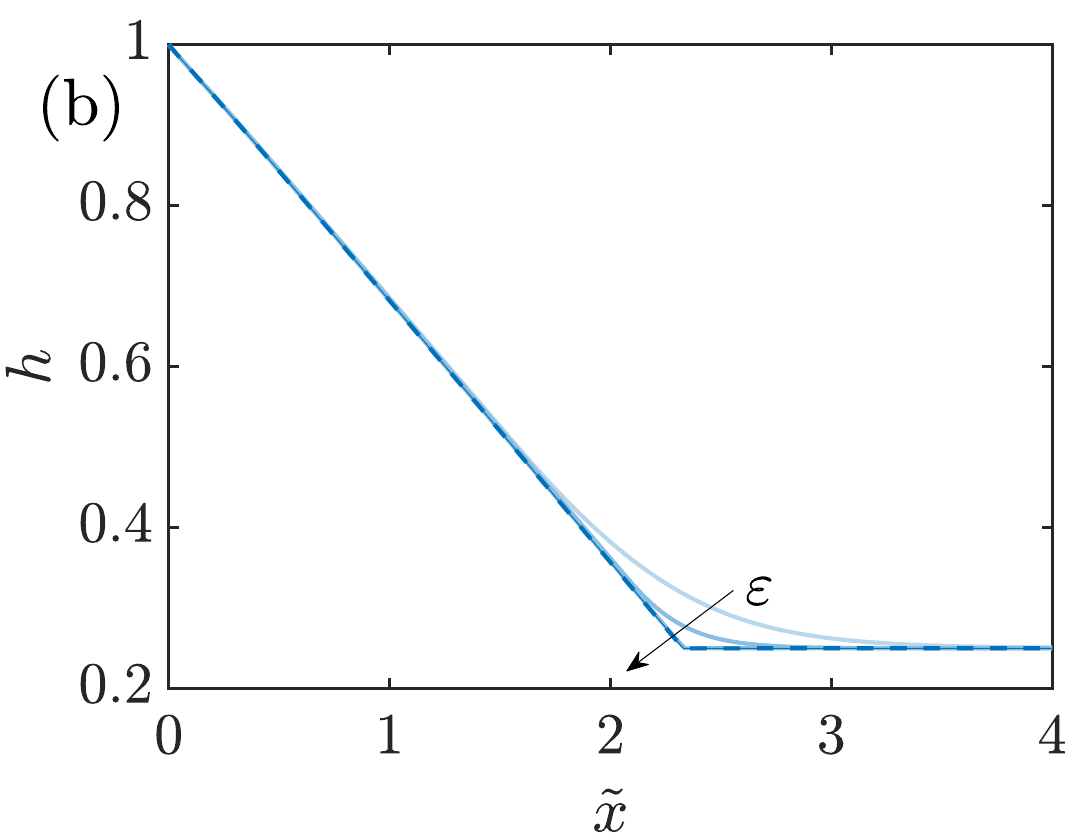}
    \includegraphics[width=0.295\linewidth]{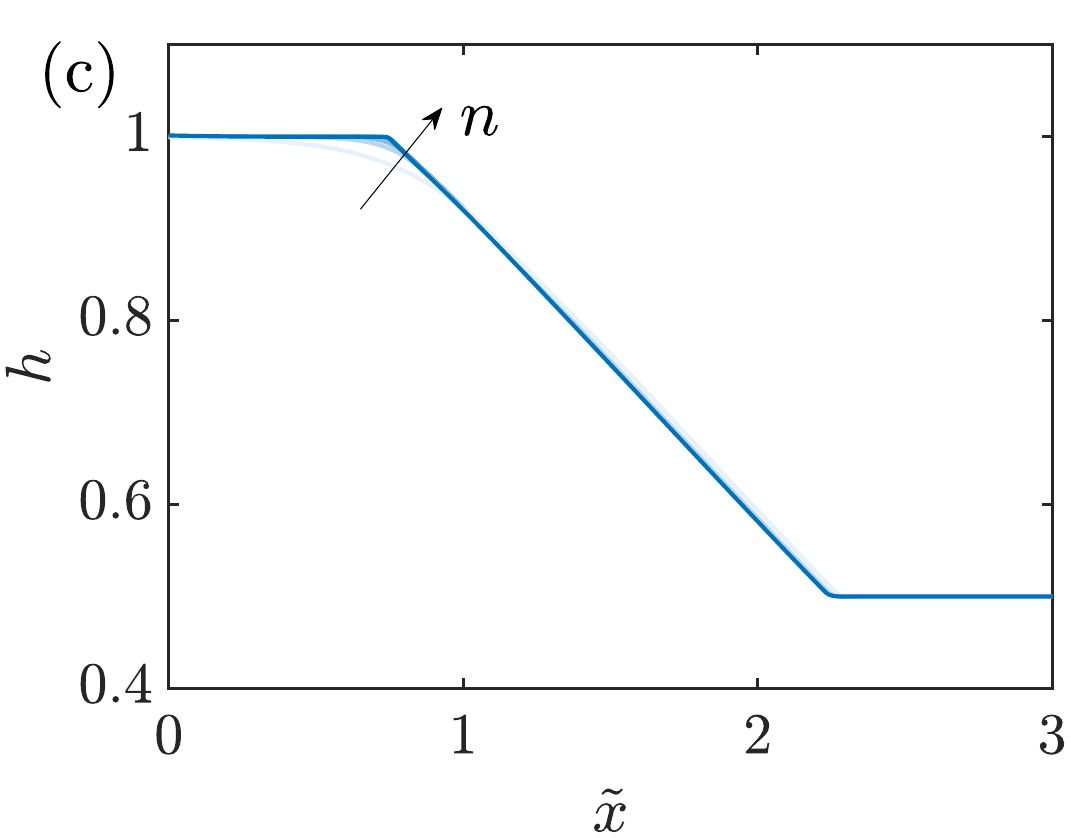}
    \caption{(a,b) Illustration of the convergence of numerical solutions with improved regularisation parameter for the wall friction, $\varepsilon$, showing (a) the solid velocity at the base, and (b) the height of the layer as  $\varepsilon$ is reduced in the direction of the arrows (light to dark blue: $\varepsilon = 1, 0.5, 0.1$ and $0.01$). (c) Convergence of the height of the layer with varying choose of smoothing function $\chi = h^n$, for $n = 10$, 25, 50, 100, 200, and 400 (light to dark blue, arrow in the direction of increasing $n$).
    Simulations are performed $\mu_s = 0.32$, $\mu_2 = 0.7$, $\mu_w = 0.55$, $K = 10^{-5}$ $\theta = 10^{-5}$}
    \label{fig:regularisation}
\end{figure}

The model presented here contains regularisation of the shear stress on the bottom and top (when the gap is bridged) of the channel, as introduced in \S\ref{sec:GranularFriction}. Figure~\ref{fig:regularisation}(a,b) illustrates the convergence of the numerical solution of our model with the regularisation parameter $\varepsilon$. Panel (b) shows that as $\varepsilon$ decreases the difference in the predicted height profile becomes indistinguishable between the two lowest values of $\varepsilon$ shown here, which are still at least an order of magnitude larger than the value we use throughout this paper ($\varepsilon = 10^{-3}$). 

We also use a regularisation method to describe the transition between bridged and un-bridged grains, as discussed after (\ref{eq:ConservationOfMass}). Here we cap the potentially diverging shear stress coming from the upper fluid, and we also smooth out the jump between bridged and unbridged fluid using a smoothed indicator function $\chi = h^n$. We find that changing the value of the cap on the shear stress has no discernible impact on the results. For the smoothed indicator function, figure~\ref{fig:regularisation}(c) shows results for different choices of the parameter $n$. At this scale, differences in $h$ are negligible for  $n$ larger than $100$; we use $n=200$ in all our simulations here.

\section{Influence of side walls} \label{sec:SideWalls}

Here we consider the effect of the side walls of the tube, which are located at $y = \pm W/2$, by developing a heuristic model to account for some side-wall friction. We still consider the velocity in the $y$ and $z$ directions to be negligible such that $\vec{v}_s = u_s \vec{e}_x$. As such, equation \eqref{eq:abc123} becomes
\begin{align}
	\frac{\p \sigma'_{xy}}{\p y} + \frac{\p \sigma'_{xz}}{\p z} = \frac{\p}{\p x} (p_s + p_f) =  \frac{\p P_f}{\p x} + \frac{\p P_s}{\p x} + \phi_s^* \Delta \rho g \frac{\p h}{\p x}, \label{eq:sigyz}
\end{align}
recalling $p_s = P_s + \Delta \rho g \phi_s^* (h-z)$ and $p_f = P_f + \rho_f g (b-z)$ as before. Taking the horizontal ($y$-direction) average of (\ref{eq:sigyz}) and assuming that any variation in the vertical shear stress and the pressures in the $y$ direction is sufficiently weak that the averages can be approximated by their local values, we arrive at
\begin{align}
	 \frac{\p {\sigma'}_{xz}}{\p z}& \approx \frac{\p {P_f}}{\p x} + \frac{\p {P_s}}{\p x} + \phi_s^* \Delta \rho g \frac{\p h}{\p x} + \frac{ \sigma_{xy}'(y = -b/2) - \sigma_{xy}'(y = b/2) }{W}, \label{eq:ShearStressSide}
\end{align} 
Analogous to the friction imposed on the top and bottom plates as described in \S \ref{sec:ModelDevelopment}, we impose a Columb friction condition on the side walls
\begin{align}
	u_s = 0  \qquad &\textrm{if } \sigma_{xy}'(y = \pm W/2) < \mu_w p_s, \label{eq:SideWallFriction1} \\
	u_s > 0  \qquad &\textrm{if } \sigma_{xy}'(y = \pm W/2) = \mu_w p_s, \label{eq:SideWallFriction2}
\end{align}
We regularise equations \eqref{eq:SideWallFriction1} and \eqref{eq:SideWallFriction2} analogous to the bottom and top plates via equations \eqref{eq:RegularisedShearz=0} and \eqref{eq:RegularisedStress2} such that
\begin{align}
	\sigma_{xy}'(y = W/2) - \sigma_{xy}'(y = - W/2) = 2 \mu_w (P_s + \phi_s^* \Delta \rho g(h-z)) \tanh \left( \frac{u_s}{\varepsilon} \right) .
\end{align}
Integrating equation \eqref{eq:ShearStressSide} with respect to $z$ gives the expression for the shear stress
\begin{align}
	\sigma'_{xz}&= \sigma'_{xz}(z = h) - (h - z) \left(  \frac{\p P_f}{\p x} + \frac{\p P_s}{\p x} + \phi_s^* \Delta \rho g \frac{\p h}{\p x} + \frac{2 \mu_w (P_s + \phi_s^* \Delta \rho g(h-z))}{W}  \tanh \left( \frac{u_s}{\varepsilon} \right)  \right) .
\end{align}
The effect of the side walls can be neglected if $W$ is sufficiently large. Conversely, if $W$ is very small we would expect the assumption that quantities depend only weakly on $y$, made before (\ref{eq:ShearStressSide}), to break down. As before, equations \eqref{eq:CoulombMuS} and \eqref{eq:CoulombMuS2} still apply, where $\mu(I_v)$ rheology (or $\mu(I)$) is applied in the region where $|\sigma_{xz}| > \mu_s p_s$ and $\p_z u_s = 0$ otherwise.
    
\end{document}